\documentclass[useAMS,usenatbib]{mn2e}
\usepackage{epsfig}
\usepackage{color}
\usepackage{subfigure}

\newcommand{\goodgap}{\hspace{\subfigtopskip} \hspace{\subfigbottomskip}}

\title[Galaxy haloes and  Yukawa-like gravitational potentials]{Systematic biases on galaxy haloes parameters from Yukawa-like gravitational potentials}
\author[V.F. Cardone \& S. Capozziello]{V.F. Cardone$^{1,2}$ and S. Capozziello$^{2,3}$\\
$^1$Dipartimento di Scienze e Tecnologie dell' Ambiente e del Territorio, Universit\`{a} degli Studi del Molise, \\
Contrada Fonte Lappone, 86090\,-\,Pesche (IS), Italy \\
$^2$Dipartimento di Scienze Fisiche, Universit\`{a} degli Studi di Napoli "Federico II",
Complesso Universitario \\ di Monte Sant'Angelo, Edificio N, via Cinthia, 80126 - Napoli, Italy \\
$^3$I.N.F.N. - Sezione di Napoli, Complesso Universitario di Monte Sant'Angelo, Edificio G, via Cinthia, 80126 - Napoli, Italy \\}

\date{Accepted xxx, Received yyy, in original form zzz}

\begin{document}
\maketitle

\begin{abstract}

A viable alternative to the dark energy as a solution of the cosmic speed up problem is represented by Extended Theories of Gravity. Should this be indeed the case, there will be an impact not only on cosmological scales, but also at any scale, from the Solar System to extragalactic ones. In particular, the gravitational potential can be different from the Newtonian one commonly adopted when computing the circular velocity fitted to spiral galaxies rotation curves. Phenomenologically modelling the modified point mass potential as the sum of a Newtonian and a Yukawa\,-\,like correction, we simulate observed rotation curves for a spiral galaxy described as the sum of an exponential disc and a Navarro\,-\,Frenk\,-\,White (NFW) dark matter halo. We then fit these curves assuming parameterized halo models (either with an inner cusp or a core) and using the Newtonian potential to estimate the theoretical rotation curve. Such a study allows us to investigate the bias on the disc and halo model parameters induced by the systematic error induced by forcing the gravity theory to be Newtonian when it is not. As a general result, we find that both the halo scale length and virial mass are significantly overestimated, while the dark matter mass fraction within the disc optical radius is typically underestimated. Moreover, should the Yukawa scale length be smaller than the disc half mass radius, then the logarithmic slope of the halo density profile would turn out to be shallower than the NFW one. Finally, cored models are able to fit quite well the simulated rotation curves, provided the disc mass is biased high in agreement with the results in literature, favoring cored haloes and maximal discs. Such results make us argue that the cusp/core controversy could actually be the outcome of an incorrect assumption about which  theory of gravity must actually be used in computing the theoretical circular velocity.

\end{abstract}


\section{Introduction}

The picture of a spatially flat universe undergoing accelerated expansion is the nowadays accepted view of our cosmo. According to the successful concordance $\Lambda$CDM model \citep{CCT92,SS00}, there are two  main ingredients in this scenario, namely dark matter (accounting for the clustering of the structures we observe) and the cosmological constant $\Lambda$ (dominating the energy budget and driving the cosmic speed up). The anisotropy spectrum of cosmic microwave background (\citealt{Boom,WMAP7}), the galaxy power spectrum with the imprinted baryon acoustic oscillations (\citealt{Eis05,P10}) and the Hubble diagram of Type Ia Supernovae (\citealt{Union,H09}) represent an incomplete list of the wide amount of data this model is able to excellently reproduce in a single scenario.

From a theoretical point of view, however, the $\Lambda$CDM is far to be satisfactory. First, the $\Lambda$ term is $\sim 120$ orders of magnitude larger than what expected from quantum field theory, while the ratio between its energy density and the matter one is coincidentally of order unity just today while it should have been much smaller or much larger than 1 over the rest of the universe history. Motivated by these unpleasing shortcomings, a plethora of models giving rise to a varying $\Lambda$\,-\,like term has been proposed mainly based on a scalar field evolving under the influence of its own self interaction potential. Needless to say, the absence of any candidate for this scalar field and the full arbitrariness in the choice of the potential are serious drawbacks of these models which therefore represent only a way to change the problems without actually solving them.

On galactic scales, the $\Lambda$ term gives a negligible contribution to the gravitational potential so that the classical Newtonian theory is usually adopted. While the evolution of the universe is driven by the cosmological constant, the formation of structure is mainly determined by the dark matter (DM) which provide the potential wells where baryons collapse to originate the visible component of galaxies. Numerical simulations allow to follow this process predicting the structure of DM haloes. Surprisingly, the theoretical expectations are not in agreement with observations on galactic scales. In particular, the density profile of DM haloes is expected to follow a double power\,-\,law with $\rho \propto x^{-\alpha} (1 + x)^{3 - \alpha}$ with $x = r/r_s$, $r_s$ a characteristic length scale and $\alpha$ giving the logarithmic slope in the inner regions. Although the debate on what the precise value of $\alpha$ is remains open, what is granted is that $\alpha$ should definitely be positive with $\alpha = 1$ for the most popular NFW model \citep{NFW}. The DM haloes are thus described by cusped models or, in other words, the logarithmic slope $\gamma = d\ln{\rho}/d\ln{r}$ never vanishes. On the contrary, rotation curves of low surface brightness galaxies (LSB) are definitely better fitted by cored models, i.e. $\gamma = 0$ in the inner regions, such as the pseudo\,-\,isothermal sphere, $\rho \propto 1/(1 + x^2)$ \citep{BT87} or the Burkert model, $\rho \propto (1 + x)^{-1} (1 + x^2)^{-1}$ \citep{B95,SB00}. As well reviewed in de Blok et al. (2010), cored models turn out to be statistically preferred over cusped ones from the fit to large samples of LSB rotation curves. Moreover, for those galaxies where a statistically acceptable fit for a cusp model is obtained, it turns out that the concentration parameter (defined later) is significantly larger than what expected from numerical simulations (see \citealt{deB10} and refs. therein for further details).

From the above picture, it is clear that, although observationally successful on cosmological scales, the $\Lambda$CDM model is far to be free of problems, especially if one also remember that there is up to now no laboratory final evidence for any of the many particles candidate to the role of DM. It is therefore worth going beyond the usual view and look at a radical revision of the underlying scheme. To this end, one can consider cosmic speed up as the first evidence of a breakdown of General Relativity (GR) as we know it. Rather than being due to a new actor on the scene, the accelerated expansion can be a consequence of gravity working in a different way than GR predicts. This consideration has attracted most interest towards braneworld\,-\,like models such as the five dimensions DGP models \citep{DGP00,LS03} or fourth order theories of gravity, where the GR Einstein\,-\,Hilbert Lagrangian is generalized by the introduction of a function $f(R)$ of the scalar curvature \citep{CAP,ODI,CF08,SF10,dFT10,CAPFARA}. It is worth stressing that, notwithstanding which is the correct modified gravity theory, should GR be incorrect on cosmological scales, one has to check whether the gravitational potential on galactic scale is. Most of modified theories indeed induce negligible changes to the gravitational potential on Solar System scale in order to pass the classical tests of gravity \citep{Will}, but this does not prevent to have significant deviations from the Newtonian potential on the much larger scale of galaxies where no direct experimental test is available.

Although modified gravity theories are investigated at cosmological scales as alternatives to dark energy, we are here interested in whether they can also impact the estimate of dark matter properties on galactic scales. Should indeed the gravitational potential be modified, the computation of the rotation curve must be done in this modified framework. On the contrary, one typically assumes that Newtonian mechanics holds and then constrains the halo model parameters by fitting the theoretical rotation curve to the observed one. We are here interested in investigating whether this incorrect procedure biases in a significant way the determination of the halo parameters and whether such a bias can explain the inconsistencies among theoretical expectations and observations. In a sense, we are wondering whether modified gravity could be a possible way to solve the cusp/core and similar problems of the DM scenario.

The plan of the paper is as follows. In Section 2, we present the modified gravitational potential used and detail the derivation of the rotation curve.  It is worth noticing that Yukawa-like corrections emerges in any analytical $f(R)$-gravity model, except $f(R)=R$, where $R$ is the Ricci scalar adopted in the Hilbert-Einstein action. Section 3 describes how we estimate  the bias induced on the halo model parameters by fitting data with the Newtonian potential while the underlying theory of gravity is non-Newtonian. The results of this analysis are discussed in Sections 4 and 5, while Section 6 summarizes our conclusions.

\section{Yukawa-like gravitational potentials}

As it is well known, Newtonian mechanics is the low energy limit of GR. Indeed, looking for a stationary and spherically symmetric solution of the Einstein equations gives the Schwarzschild metric whose $tt$ component gives to the Newtonian $1/r$ gravitational potential in the weak field limit. Modifying GR leads to modified field equations hence to a different solution and potential in the low energy limit. As such, we should first choose a modified theory of gravity to finally get the modified potential.

{\bf An interesting example for its application to cosmology is provided by $f(R)$ theories of gravity. Writing the metric in the weak field limit as

\begin{displaymath}
ds^2 = -[ 1 - 2 A(r) ]dt^2 + [1 + 2 B(r)] dr^2 + r^2 d\Omega^2 \ ,
\end{displaymath}
and using the conformal equivalence with scalar\,-\,field theories, Faulkner et al. (2007) have shown that\,:

\begin{equation}
A(r) = \tilde{A}(\tilde{r}) + \frac{\psi(\tilde{r})}{\sqrt{6} M_{pl}}
\end{equation}
where tilted quantities are evaluated in the Einstein frame (with $\tilde{r} = \chi^{-1/2} r$ and $\chi = f^{\prime}(R)$) and $\psi(\tilde{r})$ is the scalar field coupled to matter determined by\,:

\begin{equation}
\frac{1}{\tilde{r}^2} \frac{d}{d\tilde{r}} \left ( \tilde{r}^2 \frac{d\psi}{d\tilde{r}} \right ) = \frac{\partial{V_{eff}}(\psi, \tilde{r})}
{\partial{\psi}} \ .
\end{equation}
The effective potential $V_{eff}$ is determined from both the functional expression for $f(R)$ and the local mass density $\rho(r)$ as\,:

\begin{equation}
V_{eff} = V(\psi) + \chi^{-1/2} \bar{\rho}(\tilde{r})
\end{equation}
with $V(\psi)$ given by Eq.(6) in Faulkner et al. (2007) and $\bar{\rho} = \chi^{-3/2} \rho$. For the particular case of a uniform sphere of mass $M_c$ and radius $R_c$ and a quadratic potential, one finally gets\,:

\begin{displaymath}
\phi(r) \propto \frac{1}{r} \left [ 1 + (1/3) \exp{(- m_{\psi} r)} \right ]
\end{displaymath}
for the gravitational potential outside $R_c$. For the general case, depending on the choice of $f(R)$, the chamaleon effect (see, e.g., Faulkner et al. 2007 and refs. therein) can take place leading to the same above potential but with a different factor $\Delta/3$ instead of $(1/3)$, with $\Delta << 1$ depending on both the source mass and the local density which is embedded in.}

Motivated by this consideration, we will postulate that the gravitational potential generated by a point mass $m$ is\,:

\begin{equation}
\phi(r) = - \frac{G m}{(1 + \delta) r} \left [ 1 + \delta \exp{\left ( -\frac{r}{\lambda} \right )} \right ]
\label{eq: phipoint}
\end{equation}
where $(\delta, \lambda)$ depend on the parameters of the theory. It is worth noticing that a Yukawa\,-\,like correction has been invoked several times in the past. For instance, Sanders (1984) showed that the observed flat rotation curves of spiral galaxies may be well fitted by this model with no need for dark haloes provided $\delta < 0$ and
$\lambda$ is adjusted on a case\,-\,by\,-\,case basis. As discussed above, Yukawa\,-\,like corrections have been obtained, as a general feature, in the framework of $f(R)$-theories of gravity\footnote{{\bf As hinted at above, for $f(R)$ theories showing the chamaleon effect, $\delta$ should be a function of the source mass $m$. We will therefore implicitly assume that all the stars in a galaxy have the same mass. Needless to say, this is far to be true, but, as far as the value of $\delta$ does not change too much with $m$, the impact of this simplifying assumption can be neglected. Alternatively, one should introduce an average over the stellar mass function, but this latter quantity is largely unknown so that we prefer not to make any arbitrary assumption on its functional form and fully neglect the dependence of $\delta$ on $m$.}} \citep{arturo} and successfully applied to clusters of galaxies setting $\delta = 1/3$ \citep{enzo}. In general, one can relate the length scale $\lambda$ to the mass of the effective scalar field introduced by the Extended  Theory of Gravity\footnote{Referring to $f(R)$-gravity, we prefer to deal with "Extended Theories of Gravity" and not modified theories of gravity since these theories are nothing else but a straightforward extension of GR where the considered action is  a generic  function of the Ricci scalar \citep{CF08,CAPFARA}. }. The larger is the mass, the smaller will be $\lambda$ and the faster will be the exponential decay of the correction, i.e., the larger is the mass, the quicker is the recovering of the classical dynamics\footnote{Note that the factor $1/(1 + \delta)$ has been explicitly introduced in order to recover the correct Newtonian potential for $\lambda \rightarrow \infty$.}. Eq.(\ref{eq: phipoint}) then gives us the opportunity to investigate in a simple and unified way the impact of a large class of modified gravity theories, among these the Extended Theories, since other details do not have any impact on the galactic scales we are interested in.

Eq.(\ref{eq: phipoint}) is our starting point for the computation of the rotation curve of an extended system. To this end, we first remember that, in the Newtonian gravity framework, the circular velocity in the equatorial plane is given by $v_c^2(R) = R d\Phi/dR|_{z = 0}$, with $\Phi$ the total
gravitational potential. Thanks to the superposition principle and the linearity of the point mass potential on the mass $m$, this latter is computed by adding the contribution from infinitesimally small mass elements and then transforming the sum into an integral over the mass distribution. For a spherically symmetric body, one can simplify this procedure invoking the Gauss theorem to find out that the usual result $v_c(r) = GM(r)/r$ with $M(r)$ the total mass within $r$. However, because of the Yukawa\,-\,like correction, the Gauss theorem does not apply anymore and hence we must generalize the derivation of the gravitational potential\footnote{The non-validity of the Gauss theorem is not a shortcoming since, as discussed in Capozziello et al. (2007), physical conservation laws are guaranteed by the Bianchi identities that must hold in any modified theories of gravity.}. Alternatively, one can also remember that $v_c(R) = R F(R, z = 0)$ being $F$ the total gravitational force and $R$ a radial coordinate. This is the starting point adopted in Cardone et al. (2010) where a general expression has been derived for the case of a generic potential giving rise to a separable force, i.e.\,:

\begin{equation}
F_p(\mu, r) = \frac{G M_{\odot}}{r_s^2} f_{\mu}(\mu) f_{r}(\eta)
\label{eq: fpgeneral}
\end{equation}
with $\mu = m/M_{\odot}$, $\eta = r/r_s$ and $(M_{\odot}, r_s)$ the Solar mass and a characteristic length of the problem. In our case, it is $f_{\mu} = 1$ and\,:

\begin{equation}
f_r(\eta) = \left (1 + \frac{\eta}{\eta_{\lambda}} \right ) \frac{\exp{(- \eta/\eta_{\lambda})}}{(1 + \delta) \eta^2}
\label{eq: frexp}
\end{equation}
with $\eta_{\lambda} = \lambda/r_s$.
Using cylindrical coordinates $(R, \theta, z)$ and the corresponding
dimensionless variables $(\eta, \theta, \zeta)$ (with $\zeta = z/r_s$), the total force then reads\,:

\begin{eqnarray}
F({\bf r}) & = & \frac{G \rho_0 r_s}{1 + \delta} \nonumber \\
~ & \times & \int_{0}^{\infty}{\eta' d\eta' \int_{-\infty}^{\infty}{d\zeta'
\int_{0}^{\pi}{f_{r}(\Delta) \tilde{\rho}(\eta', \theta', \zeta') d\theta'}}}
\label{eq: forcetotgen}
\end{eqnarray}
with $\tilde{\rho} = \rho/\rho_0$, $\rho_0$ a reference density, and we have defined

\begin{equation}
\Delta = \left [ \eta^2 + \eta'^2 - 2 \eta \eta' \cos{(\theta - \theta')} + (\zeta - \zeta')^2 \right ]^{1/2} \ .
\label{eq: defdelta}
\end{equation}
Since we will be interested in axisymmetric systems, we can set $\tilde{\rho} = \tilde{\rho}(\eta, \zeta)$.
Moreover, the systems of interest here are spiral galaxies which we will model as the sum of an infinitesimally thin disc and a spherical halo,
so that a convenient choice for the scaling radius $r_s$ will be the disc scale length $R_d$.
Under these assumptions, the rotation curve may then be evaluated as\,:

\begin{eqnarray}
v_c^2(R) & = & \frac{G \rho_0 R_d^2 \eta}{1 + \delta} \nonumber \\
~ & \times & \int_{0}^{\infty}{\eta' d\eta' \int_{-\infty}^{\infty}{\tilde{\rho}(\eta', \zeta') d\zeta'
\int_{0}^{\pi}{f_{r}(\Delta_0) d\theta'}}}
\label{eq: rotcurvegen}
\end{eqnarray}
with

\begin{equation}
\Delta_0 = \Delta(\theta = \zeta = 0) =
\left [ \eta^2 + \eta'^2 - 2 \eta \eta' \cos{\theta'} + \zeta'^2 \right ]^{1/2} \ .
\label{eq: defdeltazero}
\end{equation}
Inserting Eq.(\ref{eq: frexp}) into Eq.(\ref{eq: rotcurvegen}) gives rise to an integral
 which has to be evaluated numerically even for the spherically symmetric case.
 It is, however, clear that the total rotation curve may be splitted in the sum of the
 usual Newtonian one and a corrective term disappearing for $\lambda \rightarrow \infty$, i.e.
 when the extended gravity has no deviations from GR on galactic scales.

\section{Estimating the bias}

Let us assume that a spiral galaxy can be modelled as the sum of an infinitesimally thin disc and a spherical halo and denote with ${\bf p}$ the halo model parameters. We can then write the total rotation curve as\,:

\begin{eqnarray}
v_c^2(R, M_d, {\bf p}_i) & = & v_{dN}^2(R, M_d) + v_{hN}^2(R, {\bf p}_i) \nonumber \\
~ & + & v_{dY}^2(R, M_d) + v_{hY}^2(R, {\bf p}_i) \nonumber
\end{eqnarray}
where $M_d$ is the disc mass, the labels $d$ and $h$ denote disc and halo related quantities, while $N$ and $Y$ refer to the Newtonian and Yukawa\,-\,like contributions. These latter terms fade away for $r >> \lambda$ so that the outer rotation curve is likely the same as the Newtonian one. On the other hand, in the inner region, the two curves may differ more or less depending on the value of $\lambda/R_d$. It is worth wondering whether such a difference may be compensated by adjusting the model parameters. That is to say, we are looking for a new set $(M_d^{\prime}, {\bf p}^{\prime})$ such that\,:

\begin{displaymath}
v_{cN}^2(R, M_d, {\bf p}) = v_{dN}^2(R, M_d^{\prime}) + v_{hN}^2(R, {\bf p}^{\prime}) \ .
\end{displaymath}
Formally, this problem could be solved explicitly writing down the above relation for ${\cal{N}} + 1$ values of $r$ with ${\cal{N}}$ the number of halo parameters and then checking that the matching between the two curves is reasonably good (if not exact) along the full radial range. Actually, such a procedure is far from being ideal since it introduces a dependence of the results on the values of $r$ chosen. Moreover, we do not need to exactly match the two curves, but only find $(M_d^{\prime}, {\bf p}^{\prime})$ in such a way that the two curves trace each other within the typical observational uncertainties.

In order to find $(M_d^{\prime}, {\bf p}^{\prime})$ taking care of typical observations, we therefore adopt the procedure sketched below\,:

\begin{itemize}

\item[i.]{Compute the theoretical circular velocity for input model parameters $(M_d, {\bf p})$ using the exact expression.} \\

\item[ii.]{Generate a simulated rotation curve by sampling the above $v_c$ and adding noise to the extracted points.} \\

\item[iii.]{Fit the simulated curve with the same disc\,+\,halo model, but using the Newtonian theory to compute $v_c(R)$.} \\

\item[iv.]{Compare the output parameters $(M_d^{\prime}, {\bf p}^{\prime})$ with the input ones $(M_d, {\bf p})$ as function of the normalized scale length $\eta_{\lambda} = \lambda/R_d$ of the modified gravity theory and other quantities of interest.} \\

\end{itemize}
In the following, we describe in more details steps i. and ii., while Sections 4 and 5 are devoted to discussing the results from a large sample of simulated curves.

\subsection{Modeling spiral galaxies}

As yet said above, we will model a spiral galaxy as the sum  of a thick disc and a spherical halo. For the disc,
we adopt a double exponential disc so that the density profile reads\,:

\begin{equation}
\rho_{d}(R, z) = \frac{M_d}{4 \pi R_d^2 z_d} \exp{\left ( - \frac{R}{R_d} - \frac{|z|}{z_d} \right )}
\label{eq: rhodisc}
\end{equation}
where $(M_d, R_d, z_d)$ are the disc mass, lengthscale and heightscale.
The Newtonian rotation curve cannot be computed analytically, but, if $z_d << R_d$ (as we will assume),
it is well approximated by the formula for the infinitesimally thin disc \citep{F70,BT87}\,:

\begin{equation}
v_{dN}^2(R) = (2 G M_d/R_d) y^2 \left [ I_0(y) K_0(y) - I_1(y) K_1(y) \right ]
\label{eq: vcdiscnewt}
\end{equation}
with $y = R/2R_d$ and $I_n(y)$, $K_n(y)$ the modified Bessel functions of order $n$ of the first and second kind, respectively.

While the observed photometry motivates the use of the exponential profile for the disc, the choice of the dark halo model is not trivial.
Numerical simulations of structure formation are typically invoked as a direct evidence favouring the use of the NFW density law or its variants.
However, the NFW model is the outcome of DM only simulations performed in a Newtonian framework, while here we are working in a modified gravity theory. In principle, one should therefore rely on the results of simulations which include both the effect of the different potential and the impact on the evolution of structure due to deviations from GR. To this end, one has first to specify which is the modified gravity theory one is considering, i.e., explicitly
write down the gravity Lagrangian. In the case of $f(R)$-gravity, \cite{S09}
have shown that, provided $f(R)$ satisfies the constraints from the Solar System, the halo density profiles of DM haloes from numerical simulations is still well approximated by the NFW model over the range of masses of interest here. Motivated by this result,
we therefore assume that the DM density profile is the NFW one\,:

\begin{figure*}
\centering
\subfigure{\includegraphics[width=5cm]{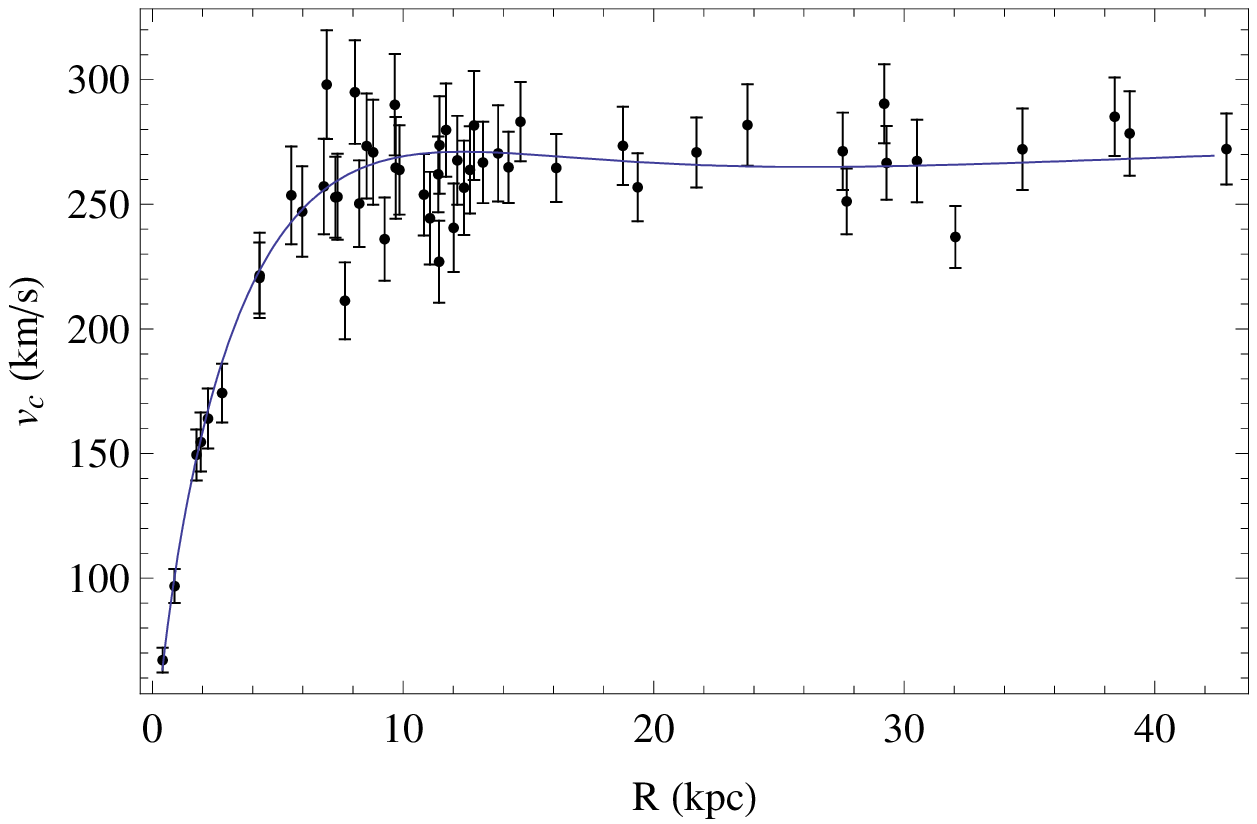}} \goodgap
\subfigure{\includegraphics[width=5cm]{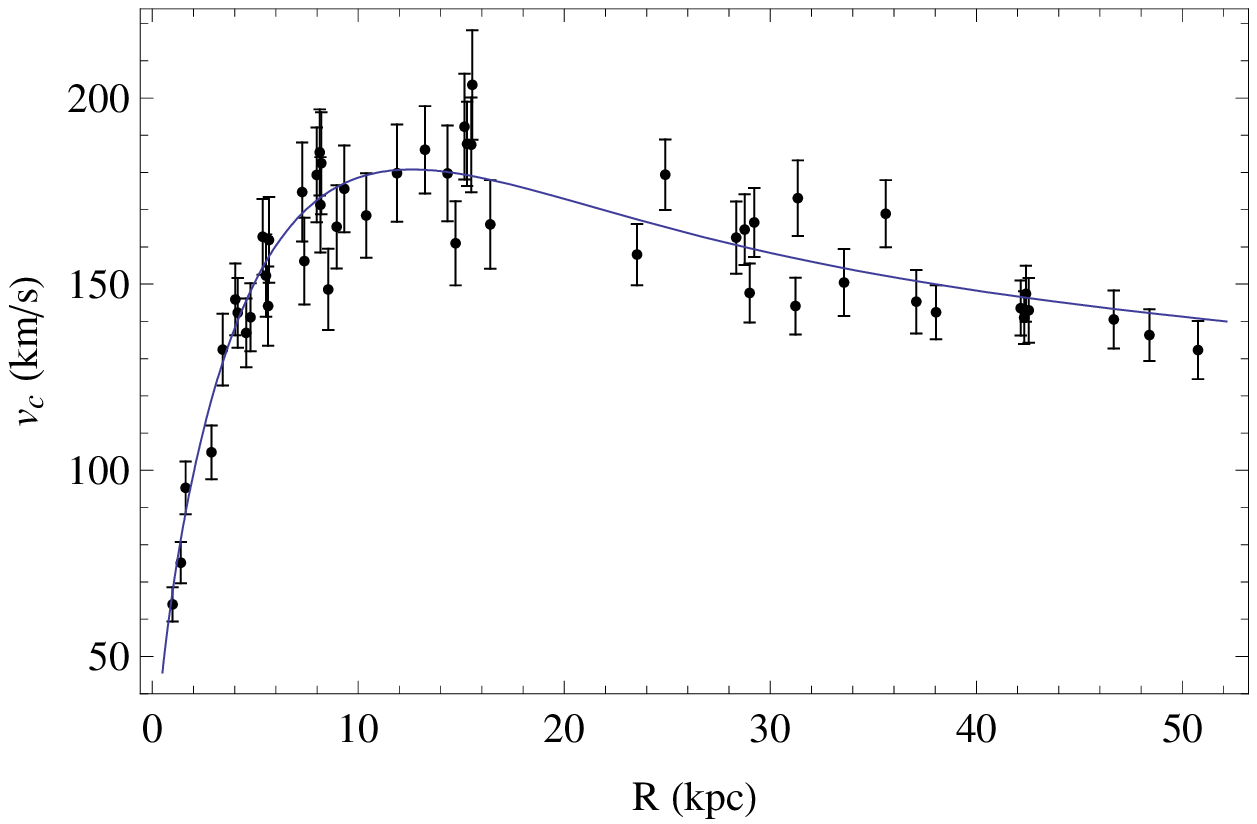}} \goodgap
\subfigure{\includegraphics[width=5cm]{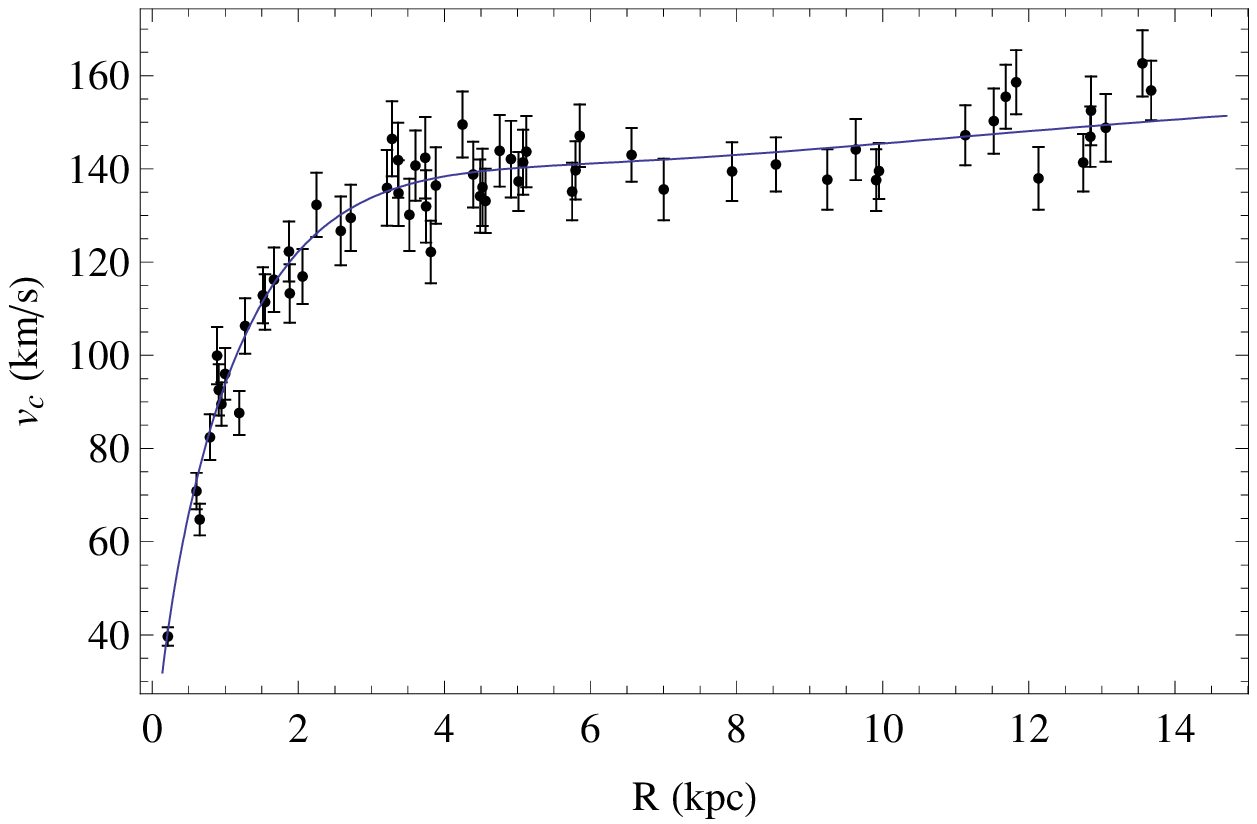}} \goodgap
\caption{Examples of simulated rotation curves with superimposed  theoretical curves. From left to right, model parameters are $(\log{M_d}, \log{M_{vir}}, c, f_{DM}, \log{\eta_{\lambda}}) = (11.15, 12.90, 10.24, 0.47, 0.36)$, $(10.90, 11.76, 14.77, 0.45, -0.92)$, $(10.04, 12.10, 13.76, 0.54, 1.11)$, while the simulation parameters are set as discussed in the text. Note that, depending on how the model parameters are set, it is possible to get rotation curves which are flat, decreasing or increasing in the outer region.}
\label{fig: egcurve}
\end{figure*}

\begin{equation}
\rho_h(r) = \frac{M_{vir}}{4 \pi R_s^3 g(R_{vir}/R_s)} \left ( \frac{r}{R_s} \right )^{-1} \left ( 1 + \frac{r}{R_s} \right )^{-2}
\label{eq: rhohalo}
\end{equation}
with

\begin{equation}
g(x) = \ln{(1 + x)} - x/(1 + x) \ .
\label{eq: defgx}
\end{equation}
In Eq.(\ref{eq: rhohalo}), $M_{vir}$ and $R_{vir}$ are the virial mass and radius. They are not independent being related by

\begin{displaymath}
R_{vir} = \left ( \frac{3 M_{vir}}{4 \pi \Delta_{th} \bar{\rho}_M} \right )^{1/3}
\end{displaymath}
with $\Delta_{th}$ the overdensity for spherical collapse and $\bar{\rho}_M = 3 H_0^2 \Omega_M/8 \pi G$ the mean matter density today. We follow \cite{BN98} for $\Delta_{th}$ and set $(\Omega_M, h) = (0.28, 0.70)$ in accordance with \cite{WMAP7}.

Because of the spherical symmetry, the Newtonian rotation curve may be easily evaluated as\,:

\begin{equation}
v_{hN}^2(r) = \frac{G M_h(r)}{r} = \frac{G M_{vir}}{R_{vir}} \frac{g(r/R_s)}{g(R_{vir}/R_s)} \ .
\label{eq: vchalonewt}
\end{equation}
While the Newtonian contributions to the rotation curve may be computed in the usual way, the Yukawa\,-\,like terms in the potential give rise to two further terms in $v_c^2(R)$ that have to be computed numerically. To this end, we must only insert Eqs.(\ref{eq: rhodisc}) and (\ref{eq: rhohalo}) into Eq.(\ref{eq: rotcurvegen}) with $f_r(\Delta_0)$ given by Eq.(\ref{eq: frexp}) subtracting the first term in parentheses. Some care must be taken in choosing the reference radius $r_s$. Since the data typically probe a limited range in $R_d$, a natural choice is to set $r_s = R_d$. However, the reference density $\rho_0$ is not the density at $r_s$. It is indeed more convenient to set $\rho_0 = \rho_d(R_d, 0)$ for the disc and $\rho_0 = \rho_h(R_s)$ for the halo. With such a choice, for the halo, the dimensionless density profile entering Eq.(\ref{eq: rotcurvegen}) is given by the $r$\,-\,dependent part of Eq.(\ref{eq: rhohalo}) provided $r/R_s$ is replaced everywhere by $\eta/\eta_s$. Finally, we remember the reader that, when computing the total rotation curve, the Newtonian terms, for both the disc and the halo, given by Eqs.(\ref{eq: vcdiscnewt}) and (\ref{eq: vchalonewt}) respectively, must be rescaled by the factor $1/(1 + \delta)$ in order to recover the classical results in the GR limit ($\lambda \rightarrow \infty$).

\subsection{Simulating the rotation curve}

In order to be useful, our approach should rely on simulated rotation curves that are as realistic as possible. By this, we mean that a.) they must refer to spiral galaxies with reasonable values of the model parameters and b.) the sampling and the noise should be the same as actual data. Point a.) is the easiest to address. As a first step, we randomly generate the disc scalelength $R_d$ and the halo virial mass from flat distributions over the ranges\,:

\begin{displaymath}
0.5 \le R_d/R_{d,MW} \le 2.0 \ \ , \ \ 11.5 \le \log{M_{vir}} \le 13.5 \ \ ,
\end{displaymath}
with $R_{d,MW} = 2.55 \ {\rm kpc}$ the disc scalelength for the Milky Way \citep{DB98,CS05}. In order to set the disc mass, we first define the halo mass fraction within the optical radius as\,:

\begin{equation}
f_{DM} = \frac{M_{h}(R_{opt})}{M_d + M_h(R_{opt})}
\label{eq: fdmopt}
\end{equation}
with $R_{opt} = 3.2 R_d$ the optical radius and we have approximated the disc mass within $R_d$ with the total disc mass.
We then randomly generate $f_{DM}$ from a flat distribution in the range $(0.9, 1.1) f_{DM,fid}$ and $f_{DM,fid} = 50\%$ (see, e.g., \cite{WPC10} and refs. therein). The halo scalelength $R_s$ is computed as $R_s = R_{vir}/c$ where the concentration $c$ is randomly generated from a Gaussian centred on\,:

\begin{equation}
c = 16.7 \left ( \frac{M_{vir}}{10^{11} \ h^{-1} \ {\rm M_{\odot}}} \right )^{-0.125}
\label{eq: cmv}
\end{equation}
and variance set to $10\%$ of the mean. Note that the above relation has been derived by \cite{Nic05} for the mass range $(0.03, 30) \times 10^{12} {\rm M_{\odot}}$ following the method detailed in \cite{B01} and updating the cosmological model. Finally, we need to set the modified potential parameters $(\delta, \lambda)$. Following the result obtained for $f(R)$-gravity \citep{enzo}, we first set $\delta = 1/3$ and run different simulations randomly generating $\log{\eta_{\lambda}} = \log{(\lambda/R_d)}$ from a flat distribution covering the wide range $(-2, 2)$. In order to explore the impact of $\delta$, we also consider the extremal case $\delta = 1.0$ thus maximizing the contribution of the Yukawa terms.

Having thus generated a realistic galaxy model, we now need a rule for sampling it and adding noise in such a way that the simulated rotation curve is similar to an observed one. A unique choice is not possible since the details of any observation depend on the instrumental setup, the observing conditions and the galaxy surface brightness. After a visual examination of rotation curves samples in literature, we have adopted the strategy summarized below.

\begin{enumerate}

\item{Take $2{\cal{N}}_{sim}$ equally spaced points $\eta_i$ in the range $(\eta_{min}, \eta_{max})$. and replace each of them with $\tilde{\eta}_i = \varepsilon_i \eta_i$ with $\varepsilon_i$ a randomly generated from the range (0.9, 1.1).} \\

\item{For each point in the sample, generate $v_{sim}(\tilde{\eta}_i)$ from a Gaussian distribution centred on the theoretical value and with variance set to $(\varepsilon_c/2) v_c(\tilde{\eta}_i)$.} \\

\item{Set the error on the $i$\,-\,th point as $\sigma_i = \delta_i \varepsilon_c v_{sim}(\tilde{\eta}_i)$ with $\delta_i$ randomly chosen in the range $(0.9, 1.1)$.} \\

\item{Generate two random numbers $(u_1, u_2)$ in the range $(0, 1)$ and take the point $i$ if $u_1 < u_2$.} \\

\end{enumerate}
Typical rotation curves are well sampled up to the optical radius $R_{opt} = 3.2 R_d$, while the sampling gets worse at larger radii. On the contrary, the percentage errors are of the same order in both regions, although it is possible that the innermost points have larger uncertainties because of deviations from ordered motions. For given input model parameters, we therefore generate two samples of points by first setting

\begin{displaymath}
({\cal{N}}_{sim}, \eta_{min}, \eta_{max}, \varepsilon_c) = (30, 0.1, 3.1, 0.25)
\end{displaymath}
and then

\begin{displaymath}
({\cal{N}}_{sim}, \eta_{min}, \eta_{max}, \varepsilon_c) = (20, 3.0, 10.0, 0.20) \ .
\end{displaymath}
We then add the two samples and rescale the errors in such a way that the standard $\chi^2$ equals 1 for the input rotation curve. Although such values are arbitrary, we have checked that the simulated rotation curves have a similar sampling and uncertainties of many dataset in literature (see Fig.\,\ref{fig: egcurve}) so that we will not try other possible combinations. In order to quantify the bias on the model parameters, we then simulate 1000 rotation curves and fit different models (as described in the following) to determine their best fit parameters. Note that we will not consider the errors on the fit quantities since they strongly depend on the uncertainties on the data points hence on the choice of the simulation parameters $({\cal{N}}_{sim}, \eta_{min}, \eta_{max}, \varepsilon_c)$. Since we are interested in a statistical analysis of the full sample rather than on a detailed study of a particular case, we prefer to limit our attention to the best fit parameters only thus avoiding to deal with the details of the simulation procedure.

\section{Bias on halo parameters}

The sample of simulated rotation curves constructed by the procedure detailed above is the starting point of our analysis. Indeed, we can now fit each curve with a given (Newtonian) model and compare the output best fit parameters with the input ones. There is, however, a preliminary caveat to be addressed. When fitting a whatever model to a given dataset, one has to put down an objective criterium to deem the model as reliable or not. It is then common practice to compute the standard $\chi^2$, i.e.

\begin{displaymath}
\chi^2 = \sum_{i = 1}^{\cal{N}}{\left [ \frac{v_c(R_i) - v_{c,th}(R_i)}{\sigma_i} \right ]^2} \ ,
\end{displaymath}
and then evaluate the quality of the fit considering the value of $\tilde{\chi}^2 = \chi^2/d.o.f.$ with $d.o.f. = {\cal{N}} - {\cal{N}}_p$ the number of degrees of freedom and ${\cal{N}}$ (${\cal{N}}_p$) the number of data points (parameters). Formally, one can then conclude that the model is a good description of the data if $\tilde{\chi}^2 \le \tilde{\chi}^{2}_{th}$ with the threshold value depending on ${\cal{N}}$. Since, for our simulated curves ${\cal{N}} \sim 50$, $\tilde{\chi}^{2}_{th} \sim 1.2$ which is highly demanding selection criterium. Actually, this formal rule rigorously applies only if the data points are uncorrelated and the measurement uncertainties are Gaussian distributed. Both these assumptions break down for our simulated curves since we have rescaled the uncertainties in such a way that $\tilde{\chi}^2 = 1$ for the input model. By a visual examination of many fitted curves, we have checked that a reasonably good fit (with no systematic deviations from the outer or inner data and rms percentage deviations smaller than $10\%$) is obtained up to $\tilde{\chi}^2 \simeq 2.5$ which we choose as our cut. We will therefore consider a model as successfully fitting the data if $\tilde{\chi}^2 \le 2.5$ and $\log{M_{vir}} \le 14$ with this latter criterium introduced to avoid unphysical solutions. However, we will also discuss the trends for the quantities of interest for a subsample obtained by using a stringent criterium, i.e. $\tilde{\chi}^2 \le 1.5$  in order to explore the impact of the threshold $\tilde{\chi}^2$ value.

\begin{figure*}
\centering
\subfigure{\includegraphics[width=7cm]{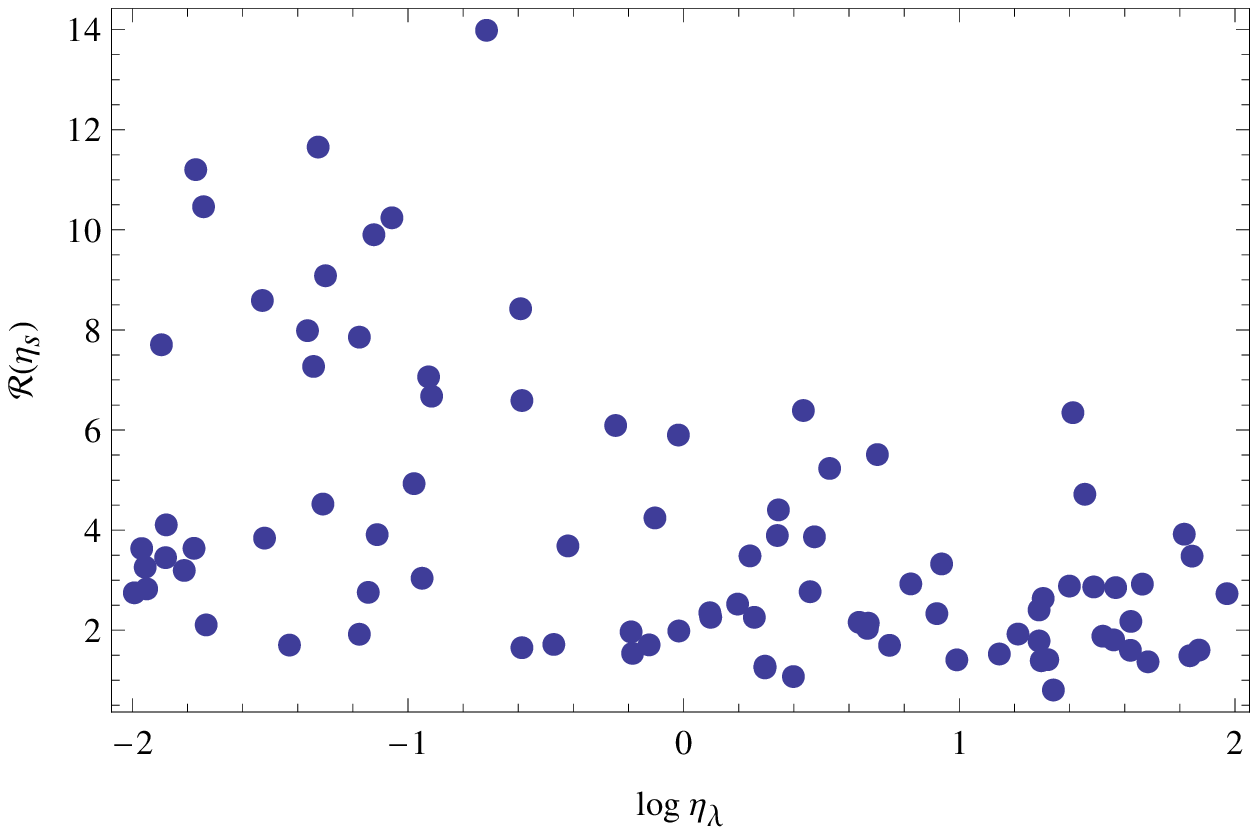}} \goodgap
\subfigure{\includegraphics[width=7cm]{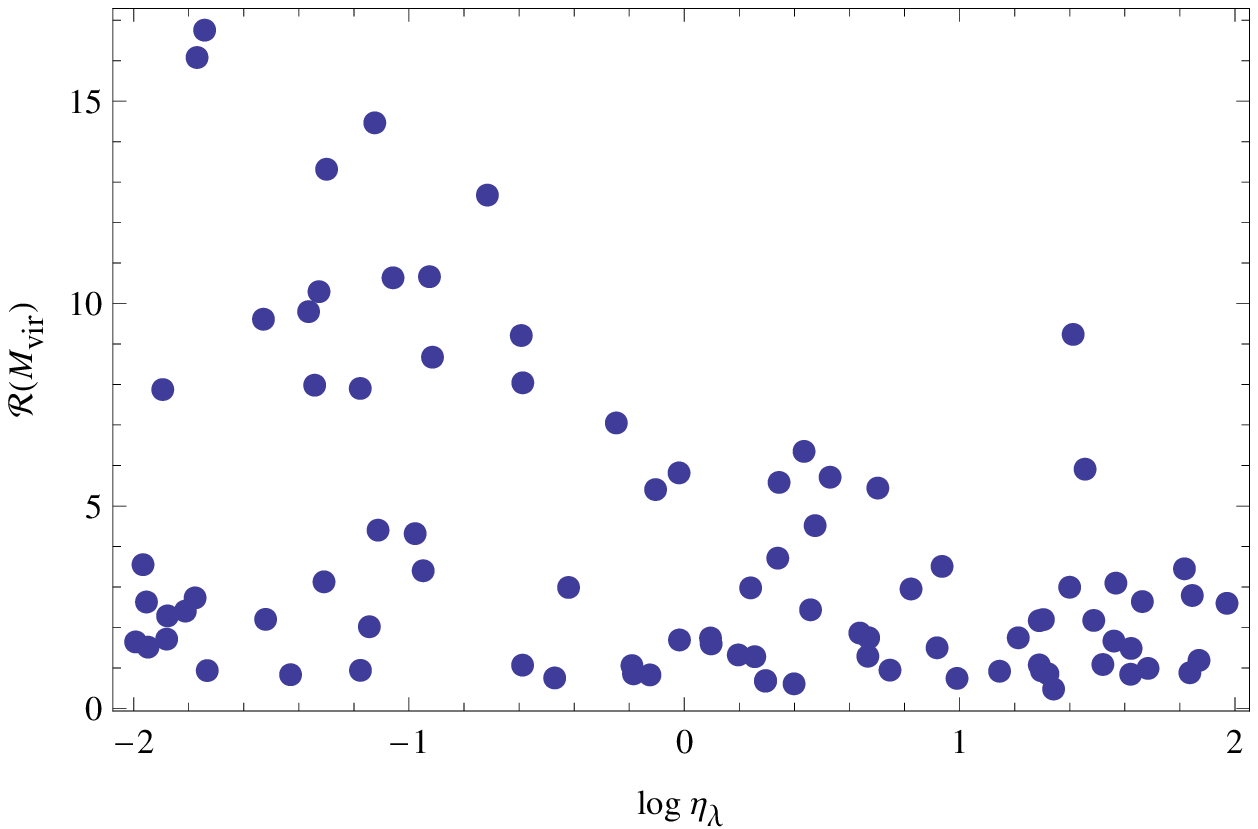}} \goodgap \\
\subfigure{\includegraphics[width=7cm]{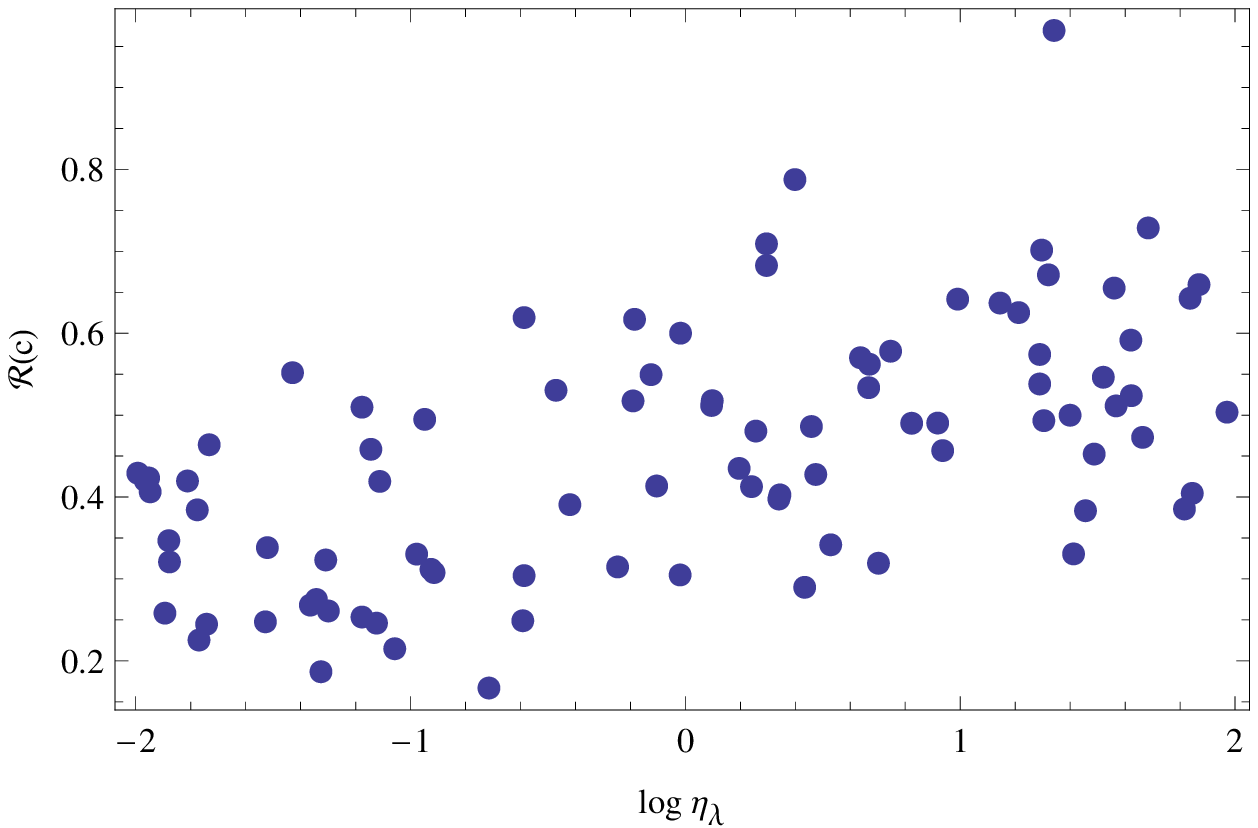}} \goodgap
\subfigure{\includegraphics[width=7cm]{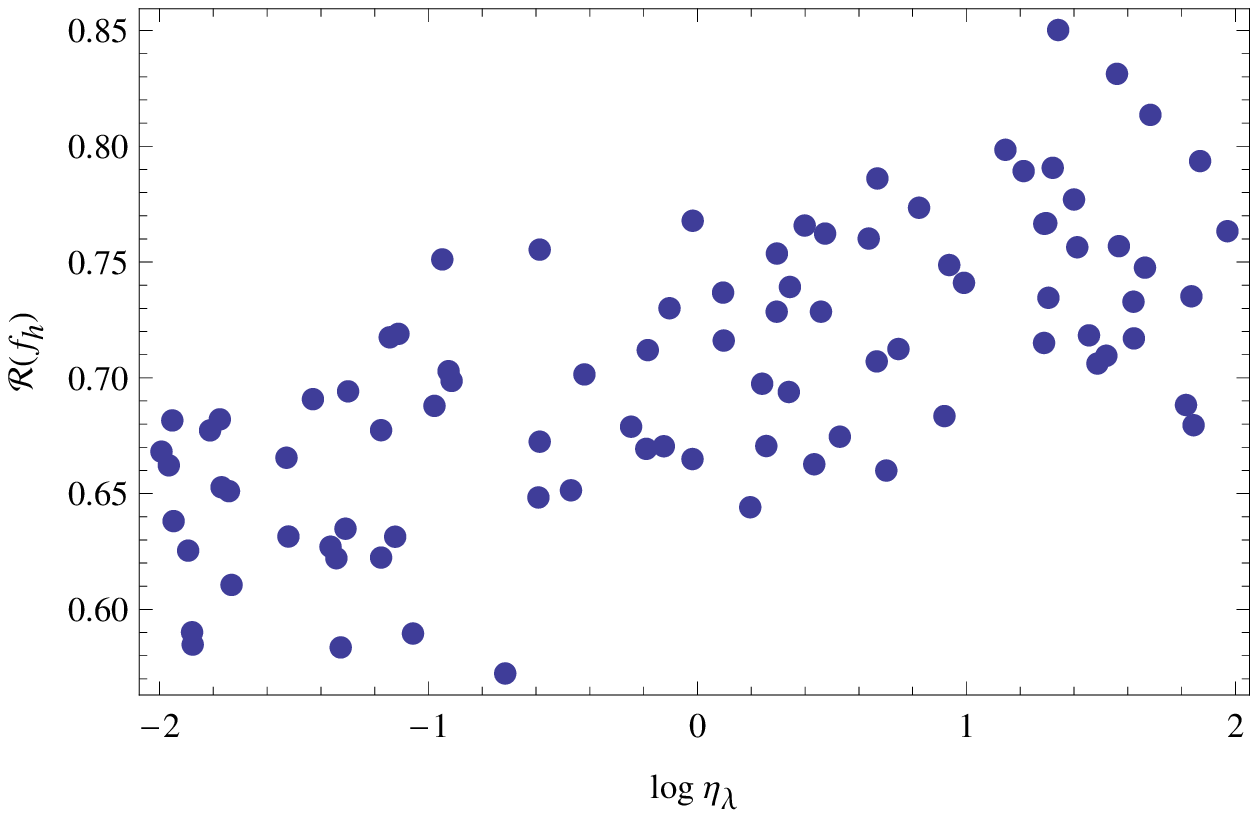}} \goodgap \\
\caption{${\cal{R}}(x)$ vs $\log{\eta_{\lambda}}$ with $x = \eta_s$ (upper left), $M_{vir}$ (upper right), $c$ (lower left), $f_{DM}$ (lower right) for the NFW model fit to the simulated rotation curves with $\delta = 1/3$. Note that only $10\%$ of the WS sample is plotted to not clutter the figure.}
\label{fig: rvsetanfw}
\end{figure*}

\begin{table*}
\caption{Bias ${\cal{R}}(x)$ on the NFW model parameters assuming the disc mass is known and $\delta = 1/3$. Columns are as follows\,:
1. parameter id; 2., 3., 4. mean $\pm$ standard deviation, median and rms values, 5., 6., 7., 8., 9. Spearman rank
correlation coefficients between  ${\cal{R}}(x)$ and input $\log{\eta_s}$, $\log{M_{vir}}$, $c$, $f_{DM}$, $\log{\eta_{\lambda}}$.
Upper half of the table is for the WS sample, while lower half for the BS one.}
\label{tab: tabnfw}
\begin{center}
\begin{tabular}{|c|c|c|c|c|c|c|c|c|}
\hline Id &  $\langle {\cal{R}} \rangle$ & ${\cal{R}}_{med}$ & ${\cal{R}}_{rms}$ & $C(\log{\eta_s}, {\cal{R}})$ & $C(\log{M_{vir}}, {\cal{R}})$ & $C(c, {\cal{R}})$ & $C(f_{DM}, {\cal{R}})$ & $C(\log{\eta_{\lambda}}, {\cal{R}})$ \\
\hline \hline

$\eta_s$ & $4.0 \pm 2.8$ & 2.9 & 4.8 & 0.31 & 0.18 & -0.17 & -0.09 & -0.47 \\

$M_{vir}$ & $3.9 \pm 3.8$ & 2.4 & 5.5 & 0.48 & 0.31 & -0.27 & -0.05 & -0.32 \\

$c$ & $0.45 \pm 0.15$ & 0.45 & 0.48 & -0.13 & -0.05 & 0.07 & 0.13 & 0.55 \\

$f_{DM}$ & $0.70 \pm 0.06$ & 0.70 & 0.71 & 0.36 & 0.30 & -0.20 & 0.39 & 0.70 \\
\hline
$\eta_s$ & $3.8 \pm 2.8$ & 2.8 & 4.7 & 0.32 & 0.24 & -0.23 & -0.07 & -0.42 \\

$M_{vir}$ & $3.6 \pm 3.6$ & 2.2 & 5.1 & 0.50 & 0.36 & -0.34 & -0.02 & -0.26 \\

$c$ & $0.47 \pm 0.15$ & 0.47 & 0.49 & -0.14 & -0.10 & 0.13 & 0.12 & 0.53 \\

$f_{DM}$ & $0.70 \pm 0.06$ & 0.70 & 0.71 & 0.39 & 0.31 & -0.18 & 0.40 & 0.71 \\
\hline
\end{tabular}
\end{center}
\end{table*}

\subsection{Navarro-Frenk-White  models}

In realistic situations, one has a set of $(R, v_c)$ data and a measurement of the galaxy surface brightness in a given band. It is then common to set the disc scalelength $R_d$ to the value inferred from photometry, while the disc mass can be inferred from the total luminosity provided an estimate of the stellar $M/L$ ratio is somewhat available (e.g., from stellar population synthesis models fitted to the galaxy colours). As a first step, we therefore assume that both $(R_d, M_d)$ are known and fit the simulated data for each rotation curve to determine the NFW model parameters\footnote{To this end, we use a straightforward Monte Carlo Markov Chain algorithm to minimize the $\chi^2$ with respect to the parameters $(\log{\eta_s}, \log{V_{vir}})$ and then infer the values of $(c, M_{vir})$ and the dark matter mass fraction $f_{DM}$. This approach is more stable than fitting directly for $(c, M_{vir})$ and allows to correctly recover the input model parameters when we fit rotation curves simulated in the Newtonian framework.} only.

\subsubsection{$\delta = 1/3$ simulated samples}

Let us first consider the results obtained setting $\delta = 1/3$ when simulating the rotation curves. As yet said above, we are here trying to fit modified gravity rotation curves using theoretical models computed in a Newtonian framework so that the first point to address is whether this is indeed possible. To this end, it is sufficient to check what is the percentage of rotation curves passing the two selection criteria quoted before. We indeed find that $90\%$ of the simulated rotation curves are fitted with $\tilde{\chi}^2 \le 2.5$ and $\log{M_{vir}} \le 14$ (which we will refer to as the {\it well fitted} sample, hereafter WS), while this fraction decreases to $77\%$ if $\tilde{\chi}^2 \le 1.5$ is demanded (defining what we will hereafter refer to as the {\it best fitted} sample, BS). Averaging over the sample values gives\,:

\begin{displaymath}
\langle \tilde{\chi}^2 \rangle = 1.27 \pm 0.24 \ (1.19 \pm 0.13) \ \ {\rm for \ WS \ (BS)}
\end{displaymath}
while for the rms of the percentage deviations $\Delta v_c/v_c = (v_{c}^{obs} - v_{c}^{fit})/v_{c}^{obs}$ we get\,:

\begin{displaymath}
\langle rms(\Delta v_c/v_c) \rangle = 6.4\% \pm 0.8\% \ (6.3\% \pm 0.7\%) \ {\rm for \ WS \ (BS)} \ .
\end{displaymath}
Motivated by the high fraction of successful fits and the low values of $\tilde{\chi}^2$ and $rms(\Delta v_c/v_c)$, we can therefore safely conclude that it is possible to fit a modified gravity rotation curve with a NFW Newtonian one obtaining both an acceptable $\tilde{\chi}^2$ value and reasonable model parameters.

Having checked that the model reasonably well fit the data, we can rely on the best fit model parameters and compare them with the input ones to estimate what is the bias induced by an incorrect assumption of the gravity law. To quantify this bias, we define ${\cal{R}}(x) = x_{fit}/x_{sim}$, i.e. the ratio between the best fit and the input values of a given quantity $x$. Needless to say, should ${\cal{R}}(x)$ be on average close to 1, we can conclude that assuming Newtonian gravity to fit modified gravity rotation curves does not induce a significant bias on the estimate of $x$. The results, summarized in Table \ref{tab: tabnfw} for both the WS and BS samples, show that such a bias is indeed quite important and only mildly depend on the selection criteria adopted (so that we will hereafter refer to the WS results only unless otherwise stated).

Table \ref{tab: tabnfw} gives some statistics on the distribution of ${\cal{R}}(x)$ for the halo parameters $(\eta_s, M_{vir})$. However, these values could be misleading since the histograms for ${\cal{R}}(\eta_s)$ and ${\cal{R}}(M_{vir})$ are strongly asymmetric with long tails towards the right. In other words, ${\cal{R}}(\eta_s)$ and ${\cal{R}}(M_{vir})$ could be much larger than their median value with ${\cal{R}}(\eta_s)$ being as high as 12 and values of ${\cal{R}}(M_{vir})$ as large as 15 so that both the halo scalelength and virial mass can be grossly overestimated. Since $c \propto R_{vir} \propto M_{vir}^{1/3}$, one could naively expect that the concentration is overestimated too. On the contrary, the distribution of the ${\cal{R}}(c)$ values is almost symmetric around its mean clearly disfavouring ${\cal{R}}(c) > 1$, i.e. the concentration is underestimated. Actually, such a result can be understood remembering that $c \propto R_s^{-1}$ so that ${\cal{R}}(c) \propto {\cal{R}}^{1/3}(M_{vir})/{\cal{R}}(\eta_s)$ finally leading to values smaller than unity. The emerging picture is therefore that of a halo having a larger mass than the input one, but also a larger scalelength. This can be qualitatively explained noting that the Yukawa\,-\,terms in the rotation curve increases the net circular velocity both in the inner and outer regions probed by the data. In order to adjust the fit in the halo dominated regions, one has to increase $M_{vir}$, but then $R_s$ has to be increased to in order to lower the density (and hence the contribution to the rotation curve) in the inner regions not to overcome the observed circular velocity. As a result, the concentration is underestimated and the same takes place for the dark matter mass fraction within the optical radius since the halo mass is now pushed outside the optical radius thus explaining why ${\cal{R}}(f_{DM})$ is smaller than unity.
As Fig.\,\ref{fig: rvsetanfw} shows, ${\cal{R}}(x)$ is correlated with $\log{\eta_{\lambda}}$ with the sign of the correlation depending on ${\cal{R}}(x)$ being typically larger or smaller than 1. Not surprisingly, the larger is the Yukawa scalelength $\lambda$, the closer is ${\cal{R}}(x)$ to 1, i.e., the smaller is the bias induced by the assumption of Newtonian gravity. However, there is a large scatter in these correlations since the relative importance of the correction also depends on the input halo parameters. Indeed, when $R_s \sim \lambda$, the halo contribute to the modified rotation curve is maximized so that the Yukawa corrections are more important and the bias of the fitted parameters is larger. As a consequence, one can therefore expect a correlation between ${\cal{R}}(x)$ and $\log{\eta_s}$ which is what we indeed find looking at the results in Table \ref{tab: tabnfw}.

\subsubsection{$\delta = 1.0$ simulated samples}

Let us now consider the results for the rotation curves sample simulated under the extreme assumption $\delta = 1.0$, i.e. when the contribution of the Yukawa term to the point mass potential (\ref{eq: phipoint}) is the same order of magnitude as the Newtonian one. Note that, different from the previous case, $\delta = 1.0$ is, as far as we know it, an unmotivated assumption, but we consider it to investigate the impact of $\delta$ on the results.

The first striking result is that the fraction of well fitted rotation curves dramatically drops to $22\%$ applying the WS selection criteria ($\tilde{\chi}^2 \le 2.5$ and $\log{M_{vir}} \le 14$), while no curves pass the cut $\tilde{\chi^2} \le 1.5$ thus showing that it is likely not possible to reproduce the simulated curves with NFW model if Newtonian gravity is assumed. Such a result can be understood noting that, for $\delta = 1.0$, the modified gravity rotation curve is much larger than the Newtonian one so that one has to greatly increase the halo virial mass to fit the outer rotation curve. In order to compensate the corresponding increase in the inner region, one should set $\eta_s$ as large as possible, but this also tend to lower the circular velocity for $R >> R_s$. Finding a compromise between these two opposite trends is quite difficult thus leading to unacceptably large $\tilde{\chi}^2$ values.

\begin{figure*}
\centering
\subfigure{\includegraphics[width=7cm]{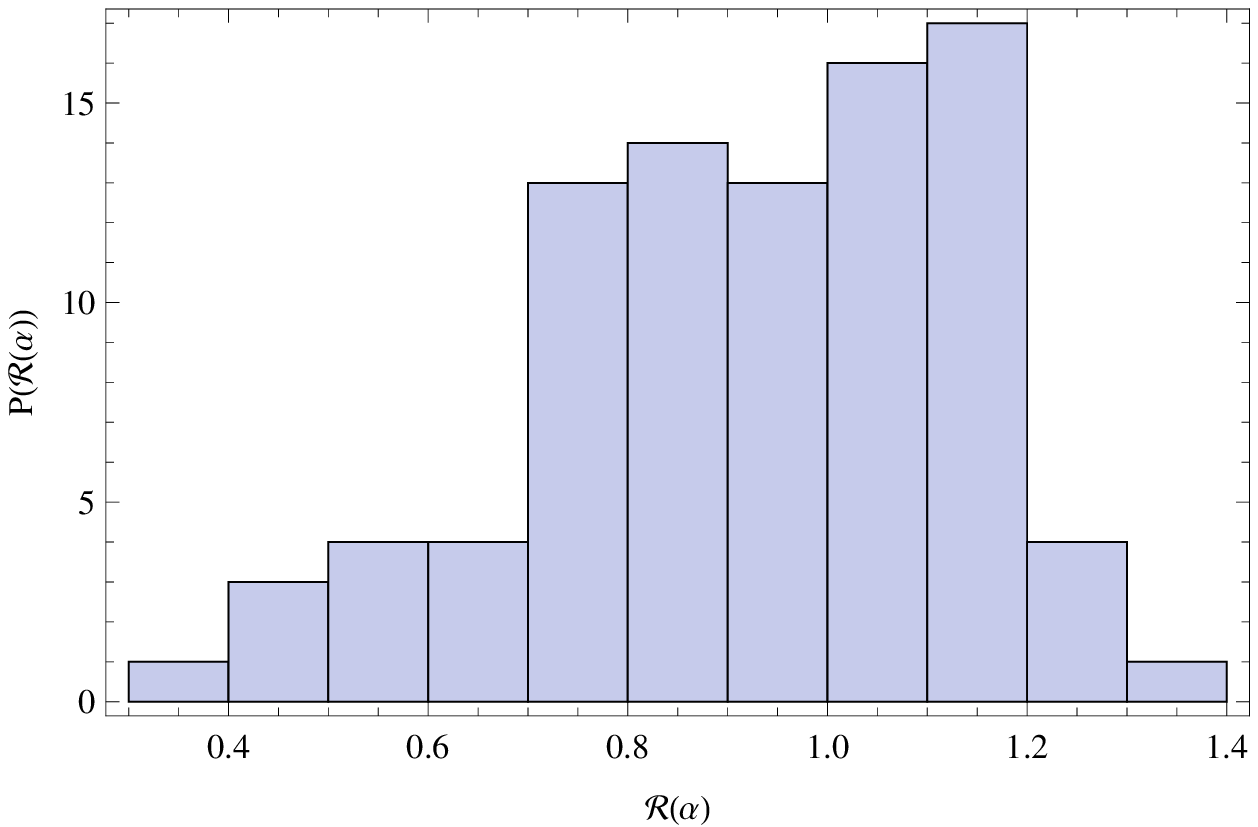}} \goodgap
\subfigure{\includegraphics[width=7cm]{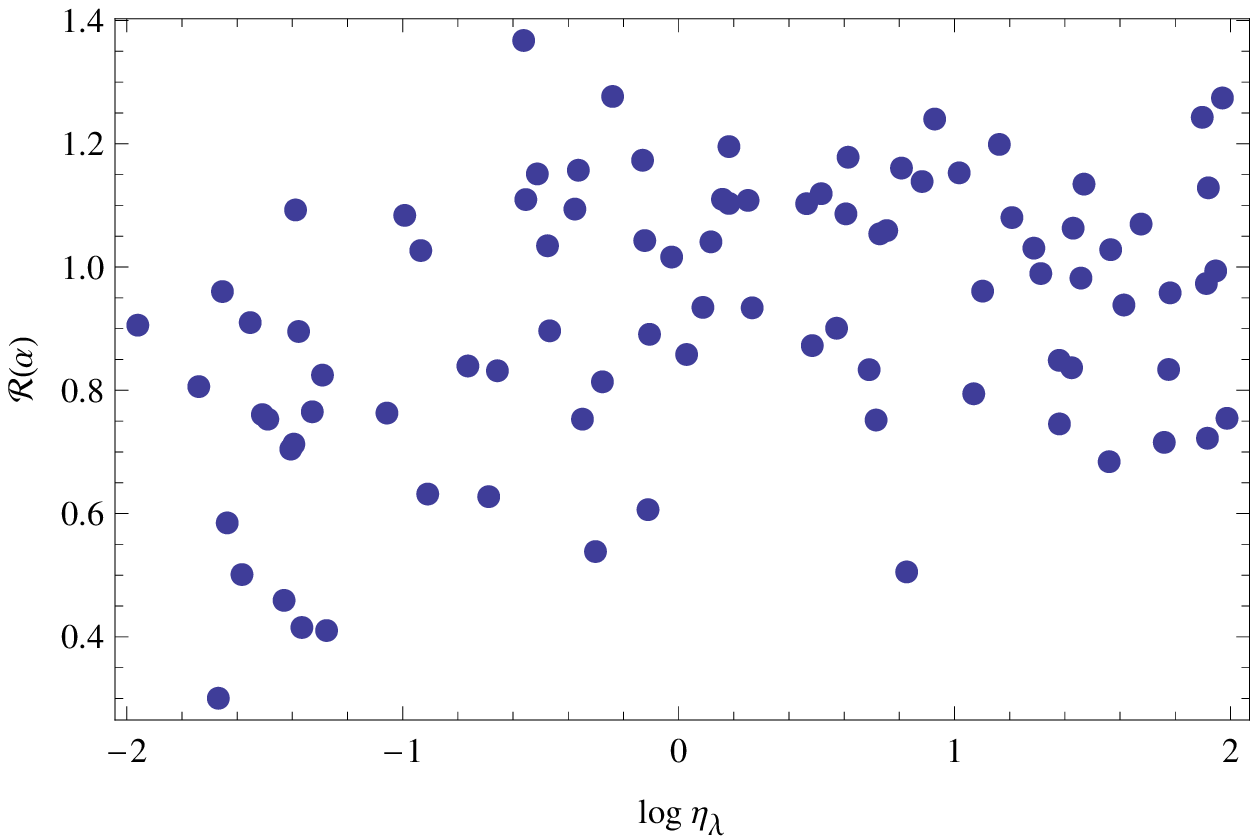}} \goodgap \\
\caption{Histogram of ${\cal{R}}(\alpha)$ values and its dependence on $\log{\eta_{\lambda}}$ for the gNFW model (with disc mass set to the input value) fit to the simulated rotation curves with $\delta = 1/3$. Note that only $10\%$ of the WS sample is plotted to not clutter the figure.}
\label{fig: gnfwplots}
\end{figure*}

It is worth investigating what is the bias induced on the halo model parameters for the successfully fitted cases. The trends of the bias parameters ${\cal{R}}(x)$ are the same as in the previous case, but the typical values are now much larger. In particular, ${\cal{R}}(\eta_s)$ can be as high as 40 thus leading to grossly underestimates of both the concentration $c$ and the dark matter mass fraction $f_{DM}$. Not surprisingly, the only way to reduce such large biases is to increase $\log{\eta_{\lambda}}$ to values larger than our upper limit $\log{\eta_{\lambda}} = 2$. However, in this case, the modified gravity potential differs from the Newtonian one only on group and cluster scales where the rotation curve are no more measured so that our analysis becomes meaningless.

\begin{table*}
\caption{As in Table \ref{tab: tabnfw}  for the gNFW with fixed disc mass case.}
\label{tab: tabgnfw}
\begin{center}
\begin{tabular}{|c|c|c|c|c|c|c|c|c|}
\hline Id &  $\langle {\cal{R}} \rangle$ & ${\cal{R}}_{med}$ & ${\cal{R}}_{rms}$ & $C(\log{\eta_s}, {\cal{R}})$ & $C(\log{M_{vir}}, {\cal{R}})$ & $C(c, {\cal{R}})$ & $C(f_{DM}, {\cal{R}})$ & $C(\log{\eta_{\lambda}}, {\cal{R}})$ \\
\hline \hline

$\alpha$ & $0.92 \pm 0.22$ & 0.95 & 0.95 & -0.56 & -0.58 & 0.37 & 0.09 & 0.32 \\

$\eta_s$ & $2.8 \pm 2.6$ & 2.0 & 3.8 & -0.11 & -0.06 & 0.0 & -0.05 & -0.06 \\

$M_{vir}$ & $2.4 \pm 2.6$ & 1.4 & 3.5 & 0.43 & 0.39 & -0.39 & 0.03 & -0.18 \\

$c$ & $0.56 \pm 0.18$ & 0.53 & 0.59 & 0.15 & 0.08 & -0.03 & -0.03 & 0.16 \\

$f_{DM}$ & $0.70 \pm 0.06$ & 0.71 & 0.71 & -0.06 & -0.18 & 0.04 & 0.40 & 0.70 \\
\hline

$\alpha$ & $0.93 \pm 0.22$ & 0.96 & 0.95 & -0.56 & -0.50 & 0.39 & 0.14 & 0.31 \\

$\eta_s$ & $2.8 \pm 2.5$ & 2.1 & 3.7 & -0.08 & -0.07 & 0.0 & 0.11 & -0.07 \\

$M_{vir}$ & $2.4 \pm 2.6$ & 1.4 & 2.7 & 0.42 & 0.36 & -0.37 & 0.05 & -0.17 \\

$c$ & $0.55 \pm 0.18$ & 0.53 & 0.58 & 0.12 & 0.08 & -0.03 & -0.09 & 0.18 \\

$f_{DM}$ & $0.70 \pm 0.06$ & 0.70 & 0.70 & -0.05 & 0.18 & 0.03 & 0.42 & 0.69 \\
\hline
\end{tabular}
\end{center}
\end{table*}
\subsection{Generalized Navarro-Frenk-White  models}

Up to now, we have considered the NFW profile for the halo model, but such a choice does not allow us to investigate whether the mismatch between modified and Newtonian gravity can have an impact on the inner logarithmic slope. In order to explore this issue, we therefore consider the generalized NFW model (gNFW, Jing \& Suto 2000)\,:

\begin{equation}
\rho_h(r) = \frac{M_{vir}}{4 \pi R_s^3 h(R_{vir}/R_s)} \left ( \frac{r}{R_s} \right )^{-\alpha} \left ( 1 + \frac{r}{R_s} \right )^{-(3 - \alpha)}
\label{eq: rhohalognfw}
\end{equation}
with

\begin{equation}
h(x) = \int_{0}^{x}{\frac{\xi^{2 - \alpha} d\xi}{(1 + \xi)^{3 - \alpha}}} \ .
\label{eq: defgxgnfw}
\end{equation}
The parameter $\alpha$ depends on the behaviour of the rotation curve in the inner regions where the disc contribute can be larger than the halo one. It is therefore worth investigating which is the impact of the uncertainties on $M_d$ in order to see how the bias on $\alpha$ depend on it.

\begin{figure*}
\centering
\subfigure{\includegraphics[width=7cm]{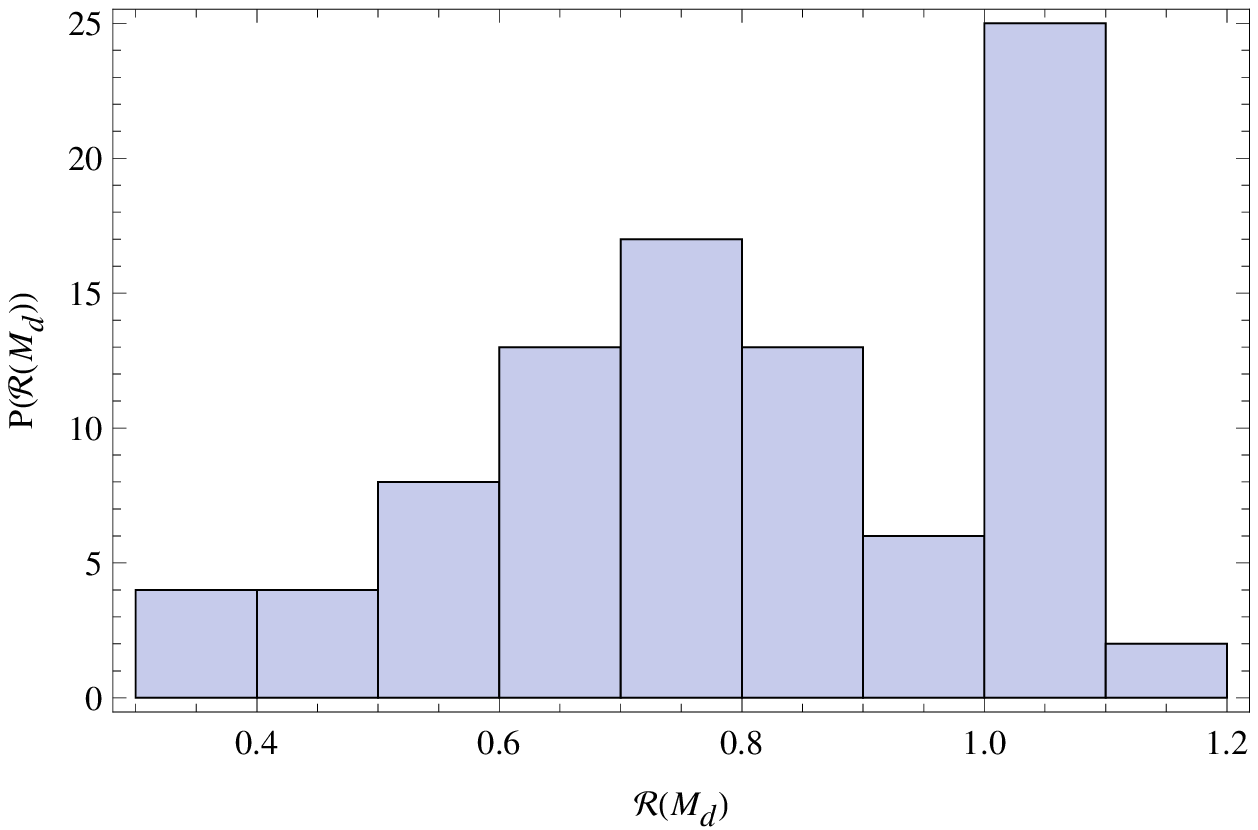}} \goodgap
\subfigure{\includegraphics[width=7cm]{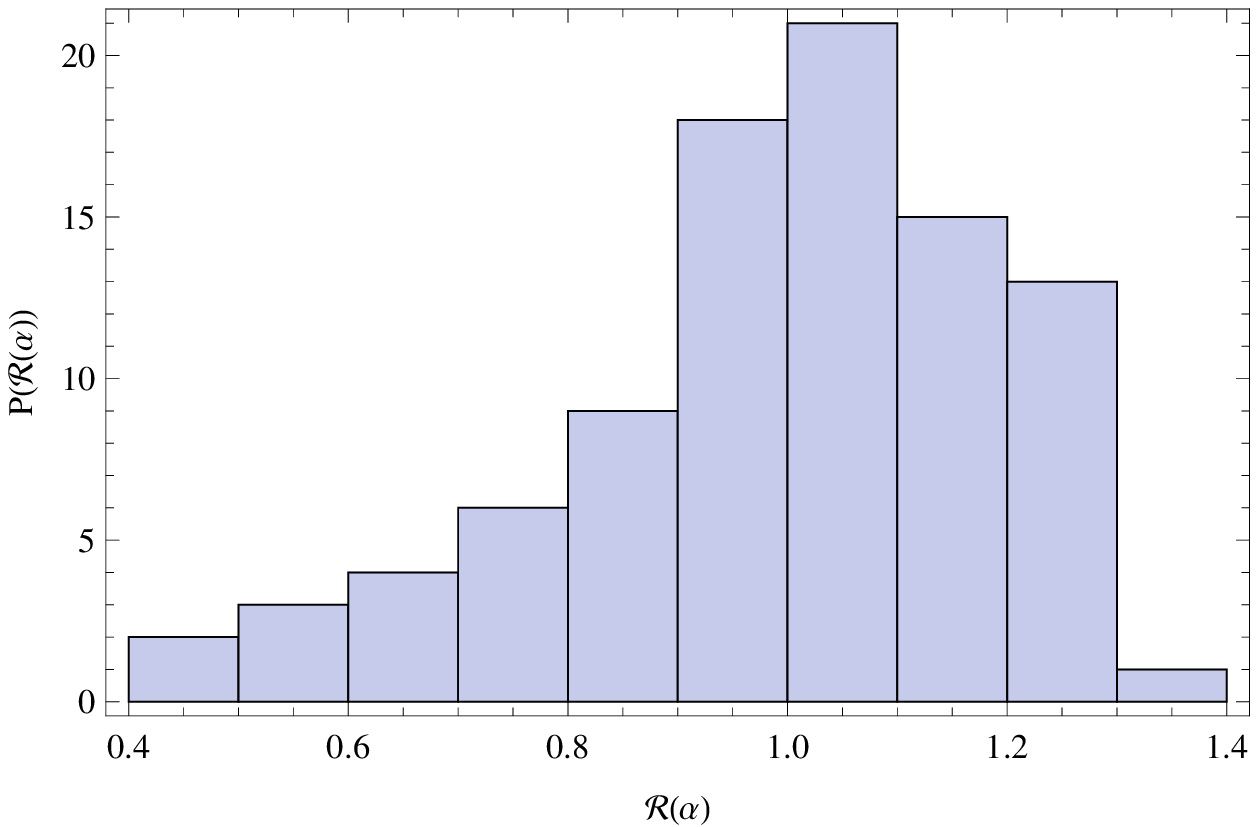}} \goodgap \\
\caption{Histograms of ${\cal{R}}(x)$ with $x = M_d$ (left) and $x = \alpha$ (right) for the gNFW model with free disc mass fits and $\delta = 1/3$.}
\label{fig: gnfwdiscplots}
\end{figure*}

\begin{table*}
\caption{As in Table \ref{tab: tabgnfw}  for the gNFW with free disc mass case.}
\label{tab: tabgnfwdisc}
\begin{center}
\begin{tabular}{|c|c|c|c|c|c|c|c|c|c|}
\hline Id &  $\langle {\cal{R}} \rangle$ & ${\cal{R}}_{med}$ & ${\cal{R}}_{rms}$ & $C(\log{M_d}, {\cal{R}})$ & $C(\log{\eta_s}, {\cal{R}})$ & $C(\log{M_{vir}}, {\cal{R}})$ & $C(c, {\cal{R}})$ & $C(f_{DM}, {\cal{R}})$ & $C(\log{\eta_{\lambda}}, {\cal{R}})$ \\
\hline \hline

$M_d$ & $0.79 \pm 0.19$ & 0.80 & 0.82 & 0.0 & 0.14 & 0.12 & -0.07 & -0.02 & 0.09 \\

$\alpha$ & $1.00 \pm 0.20$ & 1.02 & 1.01 & -0.10 & -0.01 & -0.08 & -0.01 & 0.01 & 0.33 \\

$\eta_s$ & $1.6 \pm 1.5$ & 1.2 & 1.6 & -0.01 & 0.21 & 0.18 & -0.15 & 0.03 & 0.0 \\

$M_{vir}$ & $1.7 \pm 3.5$ & 0.8 & 3.8 & 0.06 & 0.22 & 0.23 & -0.18 & 0.02 & -0.09 \\

$c$ & $0.87 \pm 0.37$ & 0.82 & 0.95 & 0.0 & -0.22 & -0.21 & 0.16 & -0.03 & 0.07 \\

$f_{DM}$ & $0.97 \pm 0.24$ & 0.96 & 1.00 & -0.04 & -0.12 & -0.12 & 0.07 & 0.05 & 0.10 \\
\hline

$M_d$ & $0.78 \pm 0.20$ & 0.78 & 0.80 & 0.03 & 0.11 & 0.13 & -0.09 & -0.04 & 0.08 \\

$\alpha$ & $1.00 \pm 0.20$ & 1.02 & 1.02 & -0.12 & 0.03 & -0.06 & -0.02 & 0.02 & 0.35 \\

$\eta_s$ & $1.6 \pm 1.6$ & 1.2 & 2.2 & 0.00 & 0.20 & 0.18 & -0.15 & 0.01 & -0.01 \\

$M_{vir}$ & $1.6 \pm 3.2$ & 0.8 & 3.5 & 0.07 & 0.21 & 0.23 & -0.18 & 0.0 & -0.12 \\

$c$ & $0.89 \pm 0.38$ & 0.84 & 0.97 & -0.02 & -0.22 & -0.22 & 0.17 & -0.01 & 0.09 \\

$f_{DM}$ & $0.99 \pm 0.24$ & 0.97 & 1.01 & -0.06 & -0.09 & -0.12 & 0.08 & 0.07 & 0.10 \\
\hline
\end{tabular}
\end{center}
\end{table*}

\subsubsection{gNFW models with known disc mass}

As a first step, we assume that $M_d$ is set to the input value and only look at the bias on the parameters $(\alpha, c, M_{vir})$. As a preliminary remark, it is mandatory explaining how the concentration is defined for gNFW models. To this end, one may simply note that, for the NFW density profile, $R_s$ is the radius at which the logarithmic slope equals the isothermal value $\gamma = -2$. Therefore, the concentration for the gNFW model may be easily defined as\,:

\begin{equation}
c = \frac{R_{vir}}{R_{-2}} = \frac{R_{vir}}{(2 - \alpha) R_s}
\label{eq: cgnfw}
\end{equation}
with $\gamma(R_{-2}) = -2$. Note that, for $\alpha = 1$, the gNFW model reduces to the NFW one and Eq.(\ref{eq: cgnfw}) gives the standard definition for the concentration of NFW haloes.

Applying the selection criteria used above to the sample with $\delta = 1/3$, we find that the WS sample contains now $90\%$ of the simulated curves, while this fraction lowers to $82\%$ for the BS sample. Averaging over the samples gives\,:

\begin{displaymath}
\langle \tilde{\chi}^2 \rangle = 1.28 \pm 0.19 \ (1.23 \pm 0.12) \ \ {\rm for \ WS \ (BS)} \ ,
\end{displaymath}

\begin{displaymath}
\langle rms(\Delta v_c/v_c) \rangle = 6.4\% \pm 0.8\% \ (6.3\% \pm 0.8\%) \ {\rm for \ WS \ (BS)} \ .
\end{displaymath}
These values are almost identical to those obtained for the fits of the NFW model described before and allows us to confidently conclude that it is possible to reproduce the simulated rotation curves with a gNFW model and Newtonian gravity. In a sense, this is not surprising since the gNFW model has one more parameter than the NFW one so that there is a much larger freedom to adjust the shape of the theoretical rotation curve to fit the observed one. As a result, the percentage of well fitted rotation curves slightly increases, but the quality estimator ($\tilde{\chi}^2$ and the rms value of $\Delta v_c/v_c$) stay almost unchanged.

The bias induced on the model parameters can be still estimated considering the quantity ${\cal{R}}(x)$ defined before and summarized in Table \ref{tab: tabgnfw} for the present case\footnote{Note that, for all the simulated curves, it is $\alpha = 1$ since we have assumed a NFW model in the simulation process. Therefore, ${\cal{R}}(\alpha) = \alpha$ for the gNFW models.}. Comparing to the results in Table \ref{tab: tabnfw}, it is apparent that the bias induced on the halo parameters in common, i.e. $(\eta_s, c, M_{vir}, f_{DM})$, is almost the same both for the WS and BS samples. The same qualitative discussion can be repeated here so that we will not consider anymore this issue. It is, on the contrary, more interesting to look at the bias on $\alpha$ which asks for some caution. As can be seen from the left panel in Fig.\,\ref{fig: gnfwplots}, the distribution of $\alpha$ values is quite large and asymmetric, while the right panel is a clear evidence that its mean value strongly depend on $\log{\eta_{\lambda}}$. Indeed, should we have only fitted models with $\lambda \le R_d$, ${\cal{R}}(\alpha)$ would have been smaller than 1, i.e. the inner slope would have been underestimated and shallower models be preferred. Should the modified gravity scalelength be smaller than the typical disc one, fitting rotation curves assuming the validity of Newtonian gravity would have us led incorrectly to believe that the inner slope of the density profile is much smaller than the NFW one. Such a result, therefore, suggests that the cusp/core controversy is just a fake problem originated by an incorrect assumption of what the underlying gravity theory actually is.

As a final remark, it is worth noting that a second difference among the results in Tables \ref{tab: tabnfw} and \ref{tab: tabgnfw} is represented by the markedly different values of the Spearman correlation coefficients. Actually, this is a consequence of having added one more parameter which introduces a degeneracy between the fitted quantities hence partially washing out the correlations of ${\cal{R}}(x)$ with the input halo parameters and the modified gravity scalelength. In particular, $\alpha$ and $\log{\eta_s}$ are clearly correlated since both set the shape of the inner rotation curve. Although with a large scatter, ${\cal{R}}(\alpha)$ is typically smaller (larger) than 1 for $\log{\eta_{\lambda}} < 0$ ($> 0$) so that ${\cal{R}}(\alpha)$ actually correlates with $\log{\eta_{\lambda}}$, but not linearly thus giving rise to a small Spearman correlation coefficient (which looks for linear correlations). Since $\eta_s$ has to be adjusted according to the changes in $\alpha$, ${\cal{R}}(\eta_s)$ has to follow ${\cal{R}}(\alpha)$ thus losing its correlation with $\log{\eta_{\lambda}}$.

Such results are strengthened if we consider the simulated curves for $\delta = 1.0$, but now relying on a much smaller statistics. Indeed, the WS sample is now made out of only $20\%$ of the full set of simulated curves, while only $2\%$ of them are included in the BS sample. These fractions are larger than in the NFW fits considered before, but are still so small to make us conclude that such extreme modified gravity curves can not be reproduced in a Newtonian framework using the gNFW model. For the few surviving curves, we can repeat the same discussion above with the only caveat that now the ${\cal{R}}(x)$ values deviate more from 1 than in the $\delta = 1/3$ case. Moreover, the slope $\alpha$ can be underestimated also when $\log{\eta_{\lambda}} > 0$ depending on the input halo parameters. This is a still further evidence in favour of the suggested interpretation of the cusp/core controversy as the outcome of a systematic error in the assumed gravity theory.

\subsubsection{gNFW models with free disc mass}

Up to now, we have set the disc mass to the input value implicitly assuming that one is able to perfectly estimate this quantity from the galaxy colors and a stellar population synthesis code. Actually, this is not always the case and, on the contrary, it is usual to include $M_d$ in the set of parameters to be determined by fitting the rotation curve data. We have therefore repeated the above analysis for the gNFW model leaving $M_d$ free to be adjusted by the fit.

Considering there is one more fit parameter, it is not surprising that the fraction of successfully fitted rotation curves (for $\delta = 1/3$) increases to $92\%$ $(85\&)$ for the WS (BS) samples and also the quality of the fit is improved being\,:

\begin{displaymath}
\langle \tilde{\chi}^2 \rangle = 1.19 \pm 0.20 \ (1.14 \pm 0.13) \ \ {\rm for \ WS \ (BS)} \ ,
\end{displaymath}

\begin{displaymath}
\langle rms(\Delta v_c/v_c) \rangle = 6.1\% \pm 0.6\% \ (6.0\% \pm 0.8\%) \ {\rm for \ WS \ (BS)} \ .
\end{displaymath}
Table \ref{tab: tabgnfwdisc} summarizes the results on the bias estimator ${\cal{R}}(x)$ for the different parameters involved which allow us to draw some interesting considerations. First, we note that the halo model parameters are now better recovered than in previous cases. For instance, although the corresponding distributions have long tails up to ${\cal{R}}(x) \sim 10 - 15$, most of the values of ${\cal{R}}(\eta_s)$ and ${\cal{R}}(M_{vir})$ are smaller than 2 leading to a modal value (i.e., the most peak of the histogram) close to 1, i.e. $(\eta_s, M_{vir})$ are not biased anymore for most of the WS and BS sample curves. A similar discussion also apply to the concentration $c$ and the dark matter mass fraction $f_{DM}$ whose histograms are quite symmetric and centred close to the no bias value, ${\cal{R}}(x) = 1$. Concerning $\alpha$, one must still keep in mind that its mean value depend on the cut on $\log{\eta_{\lambda}}$ typically being ${\cal{R}}(\alpha) < 1$ (i.e., the inner slope is shallower than the input one) when $\lambda < R_d$. Note, however, that this effect is now less pronounced than before thus explaining why the distribution of ${\cal{R}}(\alpha)$ values is peaked close to 1 with a not too large width. On the contrary, the disc mass $M_d$ turns out to be biased low with $M_d$ being smaller than the input value for most of the cases. Such an unexpected result can be qualitatively understood considering that the Yukawa terms make the modified gravity rotation curve larger than the Newtonian one in the inner and outer regions. As in the case of the NFW fit, one must increase the halo virial mass to recover this additional contribution in the halo dominated regions which increases also the halo inner circular velocity. In the NFW case, one has then to increase $\eta_s$ to prevent overestimating the simulated curve, while the same effect is now accomplished by lowering the disc mass. This explains why the halo parameters turn out to be less biased than in the NFW fit case. It is, however, worth stressing that, as shown in the left panel of Fig.\,\ref{fig: gnfwdiscplots}, the ${\cal{R}}(M_d)$ distribution is actually bimodal. As a consequence, when the disc mass is not biased low, one has again to increase $\eta_s$ thus explaining the long tails towards ${\cal{R}}(\eta_s) >> 1$ values. Alternatively, one can leave $\eta_s$ unchanged, but make the density profile shallower hence giving rise to the tail towards small ${\cal{R}}(\alpha) < 1$ in the right panel of Fig.\,\ref{fig: gnfwdiscplots}. Model degeneracies also explain why the correlation coefficients among ${\cal{R}}(x)$ and the input model parameters are typically quite small. We, however, caution the reader that a small value of Spearman correlation does not necessarily imply that ${\cal{R}}(x)$ doe not depend on the corresponding quantity (see, e.g., the ${\cal{R}}(\alpha)$ dependence on $\log{\eta_{\lambda}}$ for the gNFW fits with fixed disc mass). We have therefore checked whether this is the case or not by directly looking at the ${\cal{R}}(x)$ vs $p_i$ plots (with $p_i$ the $i$\,-\,th input parameter) finding that the scatter of the points in each plot clearly indicates a complicated dependence a weak dependence on all the input parameters but $\log{\eta_{\lambda}}$.

\begin{table*}
\caption{As in Table \ref{tab: tabnfw}  for the PI model with free disc mass case.}
\label{tab: tabpidisc}
\begin{center}
\begin{tabular}{|c|c|c|c|c|c|c|c|c|c|}
\hline Id &  $\langle {\cal{R}} \rangle$ & ${\cal{R}}_{med}$ & ${\cal{R}}_{rms}$ & $C(\log{M_d}, {\cal{R}})$ & $C(\log{\eta_s}, {\cal{R}})$ & $C(\log{M_{vir}}, {\cal{R}})$ & $C(c, {\cal{R}})$ & $C(f_{DM}, {\cal{R}})$ & $C(\log{\eta_{\lambda}}, {\cal{R}})$ \\
\hline \hline

$M_d$ & $1.20 \pm 0.10$ & 1.20 & 1.20 & -0.12 & -0.26 & -0.21 & 0.09 & 0.55 & 0.48 \\

$M_{vir}$ & $5.2 \pm 7.0$ & 3.8 & 8.7 & 0.02 & 0.12 & 0.09 & -0.03 & 0.0 & 0.06 \\

$f_{DM}$ & $0.37 \pm 0.08$ & 0.38 & 0.38 & 0.08 & 0.49 & 0.38 & -0.30 & -0.06 & -0.22 \\
\hline

$M_d$ & $1.19 \pm 0.10$ & 1.19 & 1.19 & -0.13 & -0.27 & -0.19 & 0.15 & 0.57 & 0.44 \\

$M_{vir}$ & $5.4 \pm 8.9$ & 3.7 & 10.0 & 0.0 & 0.06 & 0.04 & 0.0 & -0.03 & 0.09 \\

$f_{DM}$ & $0.38 \pm 0.08$ & 0.39 & 0.39 & 0.07 & 0.56 & 0.42 & -0.34 & -0.10 & -0.27 \\
\hline
\end{tabular}
\end{center}
\end{table*}

\begin{table*}
\caption{As inTable \ref{tab: tabnfw}  for the Burkert model with free disc mass case.}
\label{tab: tabburdisc}
\begin{center}
\begin{tabular}{|c|c|c|c|c|c|c|c|c|c|}
\hline Id &  $\langle {\cal{R}} \rangle$ & ${\cal{R}}_{med}$ & ${\cal{R}}_{rms}$ & $C(\log{M_d}, {\cal{R}})$ & $C(\log{\eta_s}, {\cal{R}})$ & $C(\log{M_{vir}}, {\cal{R}})$ & $C(c, {\cal{R}})$ & $C(f_{DM}, {\cal{R}})$ & $C(\log{\eta_{\lambda}}, {\cal{R}})$ \\
\hline \hline

$M_d$ & $1.14 \pm 0.11$ & 1.13 & 1.15 & 0.0 & -0.09 & -0.03 & -0.02 & 0.32 & 0.30 \\

$M_{vir}$ & $1.1 \pm 1.1$ & 0.7 & 0.8 & 0.06 & 0.13 & 0.16 & -0.15 & 0.09 & 0.13 \\

$f_{DM}$ & $0.46 \pm 0.11$ & 0.44 & 0.48 & -0.06 & -0.03 & -0.07 & 0.11 & 0.05 & -0.08 \\
\hline

$M_d$ & $1.11 \pm 0.10$ & 1.10 & 1.12 & -0.11 & 0.03 & 0.05 & 0.05 & 0.33 & 0.49 \\

$M_{vir}$ & $1.0 \pm 1.1$ & 0.6 & 1.5 & 0.05 & 0.12 & 0.14 & -0.10 & -0.10 & 0.21 \\

$f_{DM}$ & $0.48 \pm 0.11$ & 0.47 & 0.50 & -0.01 & -0.05 & -0.07 & -0.01 & 0.11 & -0.22 \\
\hline
\end{tabular}
\end{center}
\end{table*}

Finally, we have repeated the above analysis setting $\delta = 1.0$ to generate the simulated rotation curves. Differently from the previous cases considered up to now, we now find that it is possible to fit the gNFW model with free disc mass to the simulated rotation curves. In a sense, the additional freedom we have now to adjust the disc mass makes it possible to recover the inner rotation curve better than in the previous models thus leading to a successful fit for $97\%$ ($85\%$) of the galaxies using the selection criteria for sample WS (BS). The quality of the fit is also remarkably good\,:

\begin{displaymath}
\langle \tilde{\chi}^2 \rangle = 1.32 \pm 0.30 \ (1.18 \pm 0.15) \ \ {\rm for \ WS \ (BS)} \ ,
\end{displaymath}

\begin{displaymath}
\langle rms(\Delta v_c/v_c) \rangle = 6.5\% \pm 1.0\% \ (6.2\% \pm 0.7\%) \ {\rm for \ WS \ (BS)} \ .
\end{displaymath}
Concerning the values of ${\cal{R}}(x)$ quantities, they are almost the same as those in Table \ref{tab: tabgnfwdisc} for the parameters $(\eta_s, c, f_{DM})$ which are therefore almost unbiased. On the contrary, for the other parameters (and the WS sample), we get\,:

\begin{displaymath}
\langle {\cal{R}}(x) \rangle = \left \{
\begin{array}{ll}
\displaystyle{0.54 \pm 0.20} & \ \ {\rm for} \ \ x = M_d \\
~ & ~ \\
\displaystyle{1.04 \pm 0.28} & \ \ {\rm for} \ \ x = \alpha \\
~ & ~ \\
\displaystyle{0.85 \pm 1.18} & \ \ {\rm for} \ \ x = M_{vir} \\
\end{array}
\right . \ ,
\end{displaymath}

\begin{displaymath}
[{\cal{R}}(x)]_{rms} = \left \{
\begin{array}{ll}
0.58 & \ \ {\rm for} \ \ x = M_d \\
~ & ~ \\
1.08 & \ \ {\rm for} \ \ x = \alpha \\
~ & ~ \\
1.5 & \ \ {\rm for} \ \ x = M_{vir} \\
\end{array}
\right . \ .
\end{displaymath}
As it is clear from the large standard deviations and rms values, the distribution of ${\cal{R}}(M_d)$ and ${\cal{R}}(M_{vir})$ are strongly asymmetric and, as a result, the one for $\alpha$ presents a non negligible amount of points with ${\cal{R}}(\alpha) > 1$. This result can be explained noting that setting $\delta = 1.0$ maximizes the contribution of the Yukawa terms so that we now have to mimic a similar effect by adjusting the model parameters in the Newtonian fitting. One possibility is to leave $\alpha$ unchanged with respect to the input value $(\alpha = 1.0)$, but increasing the halo mass by a significant factor thus leading to large values of ${\cal{R}}(M_{vir})$. But, in this case, one has also to decrease the disc contribution to the inner rotation curve to compensate for the additional dark matter. On the contrary, one can leave almost unchanged $M_{vir}$, but redistribute the dark mass pushing it towards the inner region. To this aim, one should make the profile steeper, i.e. increase $\alpha$, while $M_d$ must be smaller to not overcome the inner circular velocity. This strategy will lead to ${\cal{R}}(\alpha) > 1$ and ${\cal{R}}(M_{vir} \sim 1$, while again it is ${\cal{R}}(M_d) < 1$. The final histograms of ${\cal{R}}(x)$ for $(M_d, \alpha, M_{vir})$ is then the outcome of a mixture of these two possible strategies whose relative contribution depends on the details of the individual rotation curves.

\section{Bias on the halo density profile}

The use of the gNFW model to fit the simulated rotation curve may also be read as a first step towards completely mismatching the halo density profile. In a sense, the gNFW model departs from the input NFW one only because of the possibly different inner logarithmic slope. It is, however, common in data analysis to use also models that differs from the NFW one both in the inner and outer regions. Moreover, the gNFW is still a cuspy model, while the cusp/core controversy comes from the observation that cored models better fit the observed rotation curves. To this end, we now drop the implicit assumption that the trial model tracks or generalizes the NFW profile and investigate whether completely different models in the Newtonian framework are in accordance with our modified gravity simulated rotation curves.

\subsection{The pseudo\,-\,isothermal sphere}

A classical example of a cored model is represented by the pseudo\,-\,isothermal (hereafter, PI) model whose density profile reads\,:

\begin{equation}
\rho_h(r) = \frac{M_{vir}}{4 \pi R_s^3 h_{pi}(R_{vir}/R_s} \left ( 1 + \frac{r}{R_s}^2 \right )^{-1}
\label{eq: rhoiso}
\end{equation}
with

\begin{equation}
h(x) = x - \arctan{x} \ .
\label{eq: defhpi}
\end{equation}
As it is clear, the PI density law approaches a constant value for $r << R_s$, while falls off as $r^{-2}$ in the outer regions. As such, the PI model is radically different from the NFW and gNFW ones so that it is interesting to see whether can fit or not the data. Following common practice, we leave the disc mass as an unknown and estimate $(M_d, R_s, M_{vir})$.

Applying the corresponding selection cuts, we find that $78\%$ of the simulated galaxies fall into the WS sample, while the BS one contains $43\%$ of the full sample. Such high fraction and the low values of both the reduced $\tilde{\chi}^2$

\begin{displaymath}
\langle \tilde{\chi}^2 \rangle = 1.49 \pm 0.35 \ (1.24 \pm 0.15) \ \ {\rm for \ WS \ (BS)} \ ,
\end{displaymath}
and of the rms of percentage residuals

\begin{displaymath}
\langle rms(\Delta v_c/v_c) \rangle = 7.0\% \pm 0.9\% \ (6.9\% \pm 0.9\%) \ {\rm for \ WS \ (BS)}
\end{displaymath}
make us confident that the Newtonian circular velocity of the PI model can indeed mimic the rotation curve predicted by modified gravity. This is in agreement with what is indeed found in the literature where galaxies rotation curves data are typically well fitted by PI haloes.

Having used now a different halo model to fit the data, it is not possible to straightforwardly compare the output parameters with the input ones. First, although we use the same symbol, the scalelength $R_s$ of the PI model is not defined by the condition $\gamma(R_s) = -2$ as for the NFW model, but rather represents the core radius. As such, it is meaningless to compare the input $\eta_s$ with the output $R_s/R_d$ since they refer to a different characteristic of the density profile. Moreover, a concentration for the PI model is not defined so that we can not compare with the input one. On the contrary, the total disc mass $M_d$, the virial mass $M_{vir}$ and the dark matter mass fraction $f_{DM}$ refer to global properties of the disc\,+\,halo model so that it makes sense to compare them with the input ones. Table \ref{tab: tabpidisc} then gives the bias on the global parameters $(M_d, M_{vir}, f_{DM})$ both the WS and BS samples (having set $\delta = 1/3$). It is clear that all these three quantities are severely biased with $(M_d, M_{vir})$ being overestimated and $f_{DM}$ grossly underestimated. Such a behaviour can be explained noting that, once the core radius has been set, the only way to increase the Newtonian rotation curve is to increase the global masses $(M_d, M_{vir})$ thus leading to asymmetric distributions both peaked in ${\cal{R}}(x) > 1$. It is worth noting that, if one ignores that the underlying gravity theory is not Newtonian, a large disc mass will be interpreted as the presence of a maximal disc which is indeed the case in many spiral galaxies \citep{PW00}. Being both $M_{vir}$ biased high, one could find the result ${\cal{R}}(f_{DM}) << 1$ an unexpected nonsense. Actually, one has first to note that also the disc mass is overestimated which automatically reduces the dark matter mass fraction. Moreover, for the PI model, the mass profile increases almost linearly, while, for the input NFW profile, $M_h(r)$ grows approximately in a logarithmic way. Being the input $R_s$ almost the same as the output core radius, we therefore get $M_{NFW}(R_{opt}) > M_{PI}(R_{opt})$ even if the virial mass of the PI model is larger than the NFW input one. As a consequence, $f_{DM}$ turns out to be underestimated as we indeed find.

As a final test, let us consider what happens when $\delta = 1.0$ is used in simulating the rotation curves. It turns out that the percentage of curves entering the WS sample reduces to $67\%$, while only a modest $12\%$ enter the BS sample. This sudden drops is actually due to the cut on the output virial mass rather than on the reduced $\chi^2$. As such, we believe that such a case can not be mimicked by a PI model under the Newtonian gravity assumption and not discuss anymore the bias on the output parameters.

\subsection{The Burkert model}

Another cored model but with a different outer profile is  the Burkert model whose density profile read \citep{B95}\,:

\begin{equation}
\rho_{h}(r) = \frac{M_{vir}}{4 \pi R_s^3 h_{B}(R_{vir}/R_s)} \left ( 1 + \frac{r}{R_s} \right )^{-1} \left ( 1 + \frac{r}{R_s} \right )^{-2}
\label{eq: rhobur}
\end{equation}
with

\begin{equation}
h_{B}(x) = \ln{(1 + x)} - \arctan{x} + (1/2) \ln{(1 + x^2)} \ .
\label{eq: defhb}
\end{equation}
Note that the Burkert model presents an inner core with radius $R_s$, but asymptotically drops off as $r^{-3}$ (hence having a finite total mass). As such, it represents a sort of compromise between the cored PI model and the cusped NFW one. Again, we will leave the disc mass free and determine the three parameters $(M_d, \eta_s, M_{vir})$ from the fit.

Setting $\delta = 1/3$, we find that $88\%$ of the galaxies fall into the WS sample, while this fraction drops to $37\%$ for the BS one. The quality of the fit may be judged from the following figures of merit\,:

\begin{displaymath}
\langle \tilde{\chi}^2 \rangle = 1.63 \pm 0.32 \ (1.31 \pm 0.15) \ \ {\rm for \ WS \ (BS)} \ ,
\end{displaymath}

\begin{displaymath}
\langle rms(\Delta v_c/v_c) \rangle = 7.3\% \pm 1.0\% \ (6.4\% \pm 0.7\%) \ {\rm for \ WS \ (BS)} \ ,
\end{displaymath}
comparable to the values obtained for the PI fits. It is worth noting that the fraction of galaxies in the WS sample is larger than for the PI model as a result of the different mass profile. Indeed, for the PI model, asymptotically $M(r) \propto r$ so that $v_c(r) \propto M(r)/r \sim const$, while for the Burkert model we have a finite total mass and hence a Keplerian fall off. As Fig.\,\ref{fig: egcurve} shows, our simulated rotation curves sample is made out of galaxies which can be flat, decreasing or increasing in the outer regions depending on the input model parameters. It is, of course, quite difficult to fit decreasing curves with the PI model which does not predict such a behaviour, although the limited radial range probed and the uncertainties allow to get a reasonable match in some cases. Models with $\rho_h \propto r^{-3}$ for $r >> R_s$ work better thus motivating the higher fraction of success of the Burkert profile.

The bias values ${\cal{R}}(x)$ are summarized in Table \ref{tab: tabburdisc} for the case with $\delta = 1/3$ and shows that, even assuming a Burkert halo, the disc mass turns out to be overestimated hence mimicking maximal disc solutions. On the contrary, the halo virial mass is only modestly biased, especially if compared to the outcome of previous fits. Although the ${\cal{R}}(M_{vir})$ distribution has a long tail to the right of the mean value, ${\cal{R}}(M_{vir}) \le 2$ for most of the galaxies in the WS sample. Since $f_{DM}$ depends on $1/M_d$ and $M_d$ is overestimated (while the dependence on $M_h(R_{opt})$ is weaker being this quantity present both at the numerator and denominator), it is then expected that $f_{DM}$ is biased low. This is indeed what we find in accordance with the similar result obtained for the PI model. Note, however, that this time the bias is smaller than for the PI model, even if still quite significant.

Finally, we have repeated the above analysis setting $\delta = 1.0$ when simulating the rotation curves. The WS sample still contains $73\%$ of the full simulated sample, but only $17\%$ of them pass the criteria for entering the BS sample. This situation is quite similar to what takes place for the PI model so that we do not discuss it here anymore.

\section{Conclusions}

General Relativity has been experimentally tested on scales up to the Solar System one so that assuming it still holds on much larger scales, such as the galactic and cosmological ones, is actually nothing else but an extrapolation. Motivated by this consideration and the difficulties in explaining the observed accelerated expansion without introducing new unknown ingredients like dark energy, a great interest has been recently devoted to modified gravity theories which have proven to work remarkably well in fitting the data and predicting the correct growth of structures. Modifying General Relativity has impact at all scales so that, provided no departures from standard results,  well established at Solar System scales, one cannot exclude a priori that the gravitational potential generated by a point mass source has not the usual Keplerian fall off, $\phi \propto 1/r$, but a weaker one. Here we have  considered the case of a Yukawa\,-\,like correction, i.e. $\phi \propto (1/r) [ 1 + \delta \exp{(-r/\lambda)}]$ where the scale length $\lambda$ is related to the effective scalar field (coupled with matter) introduced by several modified theory of gravity.  In particular, this kind of potential comes out in the weak field limit of $f(R)$-gravity. Provided $\lambda$ is much larger than the Solar System scale, the corrections to the potential can significantly boost the circular velocity for an extended system like a spiral galaxy. Assuming Newtonian gravity to compute the theoretical rotation curve to the observed one then introduces a systematic error which can bias the estimate of the galaxy parameters thus leading to misleading conclusions.

To investigate this issue, we have fitted the Newtonian circular velocity of some widely used halo models to a large sample of simulated rotation curves computed using the assumed modified potential and a disc\,+\,NFW profile. Comparing the input parameters with the best fit ones allows us to draw some interesting lessons on the consequences of a systematic error in the adopted gravity theory. As a general result, we find that, for cusped halo models, the disc mass is underestimated, while the halo scalelength and virial mass are  biased high. Since the concentration $c$ is also biased low, the $c$\,-\,$M_{vir}$ relation will have a different slope and intercept than the one we have used to generate the model based on the outcome of N\,-\,body simulations. Moreover, if left free in the fit, the inner slope of the density profile turns out to be biased low if $\lambda$ is smaller than the disc half mass radius $R_d$. It is interesting to note that halo models shallower than the NFW one in the inner regions with $(c, M_{vir})$ values not consistent with the prediction of N\,-\,body simulations and with maximal discs are a common outcome of fitting observed (not simulated) rotation curves. Such results are usually interpreted as failures of the $\Lambda$CDM model due to misleading assumptions on the dark matter particles properties (such as their being hot or cold and their interaction cross sections) or to having neglected the physics of baryons in the simulations. Our analysis point towards a different explanation considering these inconsistencies as the outcome of forcing the gravitational potential to be Newtonian when it is not. In order to test such an hypothesis, one should fit a homogenous set of well sampled and radially extended rotation curves assuming a theoretically motivated modified gravity potential (to set the strength $\delta$ of the corrective term) and a classical NFW model (since it is in agreement with N\,-\,body simulations also in fourth order theories).

Should the scalelength $\lambda$ of the modified potential be smaller than the disc one $R_d$, we have shown that forcing the potential to be Newtonian may also lead to completely mismatch not only the model parameters, but also the halo density profiles. Indeed, cored models, such as the PI and Burkert ones, turn out to be able to fit equally well the rotation curve data. We have not carried out here a case\,-\,by\,-\,case comparison among the different models considered since this will depend critically on what the actual (not the simulated) uncertainties are and on the sampling and radial extent of the data. We can however anticipate that, for most cases, the cored models will work better than the cusped ones since they are better able to increase the inner rotation curve redistributing the total dark matter mass inside the core thus leaving almost unchanged the outer rotation curve. Such a result may have deep implications on the cusp/core controversy which should then be read as an evidence of an inconsistent assumption about the correct underlying gravity theory. From an observational point of view, one could try to fit our NFW\,+\,modified potential to LSB galaxies since this dark matter dominated systems are usually considered the best examples of the cusp/core problem. Note that our simulated sample does not actually contain LSB\,-\,like galaxies since we have set the input dark matter mass fraction as $f_{DM} \sim 50\%$ and adjusted the disc parameters having in mind a Milky Way\,-\,like spiral galaxy. Should we have simulated only LSB\,-\,like systems, we expect that cored models will definitely be preferred over cusped ones thus further strengthening our conclusions.

As a general remark, we have also obtained that it is actually quite difficult (if not impossible) to fit in a satisfactory way the simulated rotation curves with Newtonian halo models if the amplitude of the modified potential is set to $\delta = 1$, i.e. when the Yukawa term in the point mass case has the same weight as the Keplerian one. This is not surprising since the boost in the circular velocity in the halo dominated regions is so large that can only be reproduced by increasing the virial mass to unacceptably large values. Although a more detailed analysis is needed, this result could be considered as an evidence against a too large deviation from the Keplerian $1/r$ scaling. Indeed, should the case $\delta = 1.0$ be realistic, then the actually observed rotation curves would resemble the simulated ones and hence we should have been unable to fit them. This is obviously not the case since a plethora of successful fits are available in literature. We can therefore argue that the case $\delta = 1$ is not realistic at all or, in other words, that a Yukawa\,-\,like deviation from the Newtonian potential must be only a subdominant correction
 to the $1/r$ scaling in the inner galaxy regions.

Although a more detailed analysis  is needed, we would finally stress that our analysis points towards a new usage of the rotation curves. Looking for inconsistencies rather than for agreement between these data and Newtonian models can indeed tell us not only whether the dark matter particles properties should be modified or not, but also whether our assumptions on the underlying theory of gravity are correct or not. Although it is likely that a definitive answer on this question could not be achieved in this way, the analysis of the rotation curves data stands out as a new tool to deal with modified gravity at scales complementary to those tested by cosmological probes. Asking for consistency among the results on such different scales could help us to select the correct law governing the dominant force of the universe.


\begin{thebibliography}{99}

\bibitem[\protect\citeauthoryear{Binney \& Tremaine }{1987}]{BT87}
Binney, J., Tremaine, S. 1987, {\it Galactic Dynamics}, Princeton University Press, Princeton (USA)

\bibitem[\protect\citeauthoryear{Bryan \& Norman }{1998}]{BN98}
Bryan, G.L., Norman, M.L. 1998, ApJ, 495, 80

\bibitem[\protect\citeauthoryear{Bullock et al. }{2001}]{B01}
Bullock, J.S., Kolatt, T.S., Sigad, Y., Somerville, R.S., Kravtsov, A.V., Klypin, A.A., Primack, J.R., Dekel, A. 2001, MNRAS, 321, 559

\bibitem[\protect\citeauthoryear{Burkert }{1995}]{B95}
Burkert, A. 1995, ApJ, 447, L25

\bibitem[\protect\citeauthoryear{Capozziello}{2002}]{CAP}
Capozziello, S. 2002, Int. Jou. Mod. Phys. D, 11, 483

\bibitem[\protect\citeauthoryear{Capozziello et al.}{2007}]{CARDO}
Capozziello, S., Cardone V.F., Troisi A. 2007, MNRAS,  375, 1423

\bibitem[\protect\citeauthoryear{Capozziello \& Francaviglia }{2008}]{CF08}
Capozziello, S., Francaviglia, M. 2008, Gen. Rel. Grav., 40, 357

\bibitem[\protect\citeauthoryear{Capozziello et al. }{2009a}]{arturo}
Capozziello, S., Stabile, A., Troisi, A. 2009a, Mod. Phys. Lett. A, 24, 659

\bibitem[\protect\citeauthoryear{Capozziello et al. }{2009b}]{enzo}
Capozziello, S., de Filippis, E., Salzano, V. 2009b, MNRAS, 394, 947

\bibitem[\protect\citeauthoryear{Capozziello \& Faraoni }{2010}]{CAPFARA}
Capozziello, S., Faraoni V.  2010, {\it Beyond Einstein Gravity: A Survey of Gravitational Theories
for Cosmology and Astrophysics}, Springer, New York.

\bibitem[\protect\citeauthoryear{Cardone \& Sereno }{2005}]{CS05}
Cardone, V.F., Sereno, M. 2005, A\&A, 438, 545

\bibitem[\protect\citeauthoryear{Cardone et al. }{2010}]{HL}
Cardone, V.F., Radicella, N., Ruggiero, M.L., Capone, M. 2010, MNRAS, 406, 1821

\bibitem[\protect\citeauthoryear{Carroll et al. }{1992}]{CCT92}
Carroll, S.M., Press, W.H., Turner, E.L. 1992, ARA\&A, 30, 499

\bibitem[\protect\citeauthoryear{de Bernardis et al. }{2000}]{Boom}
de Bernardis, P., Ade, P.A.R., Bock, J.J., Bond, J.R., Borrill, J., et al. 2000, Nat, 404, 955

\bibitem[\protect\citeauthoryear{de Blok }{2010}]{deB10}
de Blok, W.J.G. 2010, in {\it Advances in Astronomy}, pp. 1\,-\,15

\bibitem[\protect\citeauthoryear{de Felice \& Tsujikawa }{2010}]{dFT10}
de Felice, A., Tsujikawa, S. 2010, Living Reviews in Relativity, 13, 3

\bibitem[\protect\citeauthoryear{Dehnen \& Binney }{1998}]{DB98}
Dehnen, W., Binney, J. 1998, MNRAS, 294, 429

\bibitem[\protect\citeauthoryear{Dvali et al. }{2000}]{DGP00}
Dvali, G.R., Gabadadze, G., Porrati, R. 2000, Phys. Lett. B, 484, 112

\bibitem[\protect\citeauthoryear{Eisenstein et al. }{2005}]{Eis05}
Eisenstein, D.J., Zehavi, I., Hogg, D.W., Scoccimarro, R., Blanton, M.R., et al. 2005, ApJ, 633, 560

\bibitem[\protect\citeauthoryear{Faulkner et al. }{2007}]{F07}
Faulkner, T., Tegmark, M., Bunn, E.F., Mao, Y. 2007, Phys. Rev. D, 76, 0603505

\bibitem[\protect\citeauthoryear{Freeman }{1970}]{F70}
Freeman, K.C. 1970, ApJ, 160, 811

\bibitem[\protect\citeauthoryear{Hicken et al. }{2009}]{H09}
Hicken, M., Wood\,-\,Vasey, W.M., Blondin, S., Challis, P., Jha, S., Kelly, P.L., Rest, A., Kirshner, R.P. 2009, ApJ, 700, 1097

\bibitem[\protect\citeauthoryear{Jing \& Suto }{2000}]{JS00}
Jing, Y.P., Suto, Y. 2000, ApJ, 529, L69

\bibitem[\protect\citeauthoryear{Komatsu et al. }{2010}]{WMAP7}
Komatsu, E., Smith, K.M., Dunkley, J., Bennett, C.L., Gold, B., et al. 2010, preprint (arXiv\,:\,1001.4538)

\bibitem[\protect\citeauthoryear{Kowalski et al. }{2008}]{Union}
Kowalski, M., Rubin, D., Aldering, G., Agostinho, R.J., Amadon, A., et al. 2008, ApJ, 686, 749

\bibitem[\protect\citeauthoryear{Lue \& Starkman }{2003}]{LS03}
Lue, A., Starkman, G. 2003, Phys. Rev. D, 67, 064002

\bibitem[\protect\citeauthoryear{Napolitano et al. }{2005}]{Nic05}
Napolitano, N.R., Capaccioli, M., Romanowsky, A.J., Douglas, N.G., Merrifield, M.R., Kujiken, K., Arnaboldi, M., Gerhard, O., Freeman, K.C. 2005, MNRAS, 357, 691

\bibitem[\protect\citeauthoryear{Navarro et al. }{1997}]{NFW}
Navarro, J.F., Frenk, C.S., White, S.D.M. 1997, ApJ, 490, 493

\bibitem[\protect\citeauthoryear{Nojiri \& Odintsov }{2007}]{ODI}
 Nojiri, S., Odintsov, S.D. 2007, Int. J. Meth. Mod. Phys. 4, 115

\bibitem[\protect\citeauthoryear{Palunas \& Williams }{2000}]{PW00}
Palunas, P., Williams, T.B. 2000, AJ, 120, 2884

\bibitem[\protect\citeauthoryear{Percival et al. }{2010}]{P10}
Percival, W.J., Reid, B.A., Eisenstein, D.J., Bachall, N.A., Budavari, T., et al. 2010, MNRAS, 401, 2148

\bibitem[\protect\citeauthoryear{Sahni \& Starobinski }{2000}]{SS00}
Sahni, V., Starobinski, A., 2000, Int. J. Mod. Phys. D, 9, 373

\bibitem[\protect\citeauthoryear{Salucci \& Burkert }{2000}]{SB00}
Salucci, P., Burkert, A. 2000, ApJ, 537, L9

\bibitem[\protect\citeauthoryear{Sanders }{1984}]{S84}
Sanders, R.H. 1984, A\&A, 136, L21

\bibitem[\protect\citeauthoryear{Schmidt et al. }{2009}]{S09}
Schmidt, F., Lima, M., Oyaizu, H., Hu, W. 2009, Phys. Rev. D, 79, 083518

\bibitem[\protect\citeauthoryear{Sotiriou \& Faraoni }{2010}]{SF10}
Sotiriou, T.P., Faraoni, V. 2010, Rev. Mod. Phys., 82, 451

\bibitem[\protect\citeauthoryear{Will }{1993}]{Will}
Will, C.M. 1993, {\it Theory and experiment in gravitational physics}, Cambridge University Press, Cambridge (UK)

\bibitem[\protect\citeauthoryear{Williams et al. }{2010}]{WPC10}
Williams, M.J., Bureau, M., Cappellari, M. 2010, in {\it Hunting for the dark\,: the hidden side of galaxy formation}, eds. V.Debattista and C.C. Popescu, AIP Conference Proceedings, 1240, 431 (preprint arXiv\,:0912.5088)

\end{thebibliography}
\end{document}